\documentclass[quantumrep,article,accept,moreauthors,pdftex,10pt,a4paper]{Definitions/mdpi}

\pdfoutput=1

\firstpage{52}
\pubvolume{5}
\issuenum{1}
\articlenumber{5}
\pubyear{2023}
\copyrightyear{2023}
\datereceived{30 November 2022} 
\dateaccepted{6 January 2023}
\datepublished{18 January 2023}
\hreflink{https://doi.org/10.3390/quantum5010005}

\usepackage{authblk}
\usepackage{amssymb}
\usepackage{amsmath}
\usepackage{amstext}
\usepackage{graphicx}
\usepackage{amsfonts}\usepackage{mathrsfs}
\usepackage[normalem]{ulem}
\usepackage{mathptmx}
\usepackage{bm}

\DeclareSymbolFont{cmsy}{OMS}{cmsy}{m}{n}
\DeclareSymbolFontAlphabet{\mathcalus}{cmsy}

\makeatletter
\def\cleartheorem#1{%
	\expandafter\let\csname#1\endcsname\relax
	\expandafter\let\csname c@#1\endcsname\relax
}
\def\clearthms#1{ \@for\tname:=#1\do{\cleartheorem\tname} }
\makeatother

\newcommand{\tr}{\operatorname{tr}}
\newcommand{\ket}[1]{\lvert{#1} \rangle}
\newcommand{\bra}[1]{{\langle {#1}\rvert}}
\newcommand{\braket}[2]{\langle {#1} | {#2} \rangle}
\allowdisplaybreaks

\newcommand{\sA}{\mathrm{A}}
\newcommand{\sB}{\mathrm{B}}
\newcommand{\sC}{\mathrm{C}}
\newcommand{\sE}{\mathrm{E}}
\newcommand{\sF}{\mathrm{F}}

\newcommand{\sU}{\mathrm{U}}
\newcommand{\sK}{\mathrm{K}}

\newcommand{\sP}{\mathrm{P}}
\newcommand{\sR}{\mathrm{R}}
\newcommand{\sS}{\mathrm{S}}

\newcommand{\sV}{\mathrm{V}}
\newcommand{\sW}{\mathrm{W}}
\newcommand{\sX}{\mathrm{X}}
\newcommand{\sY}{\mathrm{Y}}
\newcommand{\sZ}{\mathrm{Z}}

\newcommand{\Span}{{\operatorname{Span}}}

\newcommand{\pr}{{\operatorname{Pr}}}

\Title{Simple and Rigorous Proof Method for the Security
of Practical Quantum Key Distribution in the Single-Qubit Regime
Using Mismatched Basis Measurements}

\TitleCitation{Simple and Rigorous Proof Method for the Security of Practical Quantum Key Distribution in the Single-Qubit Regime Using Mismatched Basis Measurements}

\Author{Michel Boyer $^{1}$,
Gilles Brassard $^{1,2}$ \orcidA{},
Nicolas Godbout $^{3}$ \orcidB{},
Rotem Liss $^{1,4,*}$ \orcidC{},
and St\'ephane Virally $^{3}$ \orcidD{}}

\AuthorNames{Michel Boyer, Gilles Brassard, Nicolas Godbout, Rotem Liss,
and St\'ephane Virally}

\AuthorCitation{Boyer, M.; Brassard, G.; Godbout, N.; Liss, R.; Virally, S.}

\address{$^{1}$ \quad D\'epartement IRO, Universit\'e de Montr\'eal,
2920 Chemin de la Tour, Montr\'eal (Qu\'ebec)\ \ H3T 1N8, Canada \\
$^{2}$ \quad QuSoft, CWI Amsterdam, Science Park 123, 1098 XG Amsterdam,
The Netherlands \\
$^{3}$ \quad Polytechnique Montr\'eal, Department of Engineering Physics,
2500 Chemin de Polytechnique, Montr\'eal (Qu\'ebec)\ \ H3T 1J4, Canada \\
$^{4}$ \quad ICFO---Institut de Ciencies Fotoniques,
The Barcelona Institute of Science and Technology, Av.\ Carl Friedrich Gauss 3,
08860 Castelldefels (Barcelona), Spain}

\corres{Correspondence: rotem.liss@icfo.eu}

\abstract{Quantum key distribution (QKD) protocols aim at allowing
two parties to generate a secret shared key.
While many QKD protocols have been proven unconditionally secure in theory,
practical security analyses of experimental QKD implementations
typically do not take into account all possible loopholes,
and practical devices are still not fully characterized
for obtaining tight and realistic key rates.
We present a simple method of computing secure key rates
for any practical implementation of discrete-variable QKD
(which can also apply to measurement-device-independent QKD),
initially in the single-qubit lossless regime,
and we rigorously prove its unconditional security
against any possible attack.
We hope our method becomes one of the standard tools used for
analysing, benchmarking, and standardizing
all practical realizations of QKD.}

\keyword{quantum key distribution; practical implementations;
security; proof}

\begin{document}

\clearthms{Proposition,Lemma,Corollary}
\newtheorem{Proposition}[theorem]{Proposition}
\newtheorem{Lemma}[theorem]{Lemma}
\newtheorem{Corollary}[theorem]{Corollary}

\section{\label{sec_intro}Introduction}
The purpose of quantum key distribution (QKD) is to allow two legitimate
parties, typically named Alice and Bob, to generate
an information-theoretically secure key~\cite{BB84}.
Most QKD protocols have been proven secure
even if the adversary Eve is allowed to apply any theoretical attack
allowed by the laws of quantum theory.
However, despite enormous progress in recent years, unconditional
security of practical implementations of QKD has remained elusive.

The difficulty of achieving practical security stems from the fact
that practical implementations deviate from the theoretical protocols
in many important aspects. The theoretical models of the preparation
devices, the transmitted quantum systems, the quantum channels,
and the measurement devices differ enormously
from any experimental realization, and these differences open up loopholes
and weaknesses that Eve may be able to exploit
(see, e.g.,~\cite{BLMS00,makarov10}).

Most security weaknesses of the measurement devices can be closed
using \emph{measurement-device-independent} (MDI)
QKD~\cite{mdi_qkd1,mdi_qkd2,mdi_qkd3,mdi_qkd4}.
However, MDI QKD still requires us to trust the preparation devices of
Alice and Bob, and deviations of the actually prepared quantum states
from the theoretical states still pose a significant security threat.
Alternatively, in \emph{(fully) device-independent} (DI)
QKD~\cite{di_qkd1,di_qkd2,di_qkd3},
Alice's and Bob's devices are completely uncharacterized,
and violations of Bell's inequality prove the secrecy of the final key.
This method, while promising and theoretically solid, still achieves
far worse secret key rates than standard QKD (including MDI QKD) in realistic
experimental settings~\cite{security_lo14,sec_practical20,qkd_review20};
in addition, it still requires assumptions, including the assumption
that Alice's and Bob's uncharacterized devices are never allowed to communicate
with each other or with Eve.
Therefore, while we believe that both DI QKD
and standard (especially MDI) QKD are important directions that can
lead to practical security (perhaps in different levels of security),
in this paper, we focus on standard and MDI QKD
protocols, where the most pressing practical security problem that has
no fully available solution is imperfectly generated quantum states.

We suggest a simple and systematic method for analysing source imperfections
and proving unconditional security of a large variety of QKD protocols.
Our method (similarly to the ``loss tolerant'' QKD
protocol~\cite{lt_qkd14,lt_qkd_exp15,lt_qkd_finite_key15,lt_qkd_gen19}) assumes
that the quantum source can only emit three possible quantum states
(instead of the four states used by BB84), and it uses a mismatched-basis
analysis (see, e.g.,~\cite{three_state_sec16}) for deriving the key rate
in the finite-key regime.
Our analysis method is vastly simplified and rigorous, takes into account
many subtle points that are often omitted in other security proofs,
and gives an explicit key rate formula in the finite-key regime.
We further suggest a practical step-by-step process
for analysing experimental implementations of QKD, and we verify that
the restriction to three states is indeed essential for practical security.

Our method currently applies only to the qubit regime (in
the generalized sense: namely, we require the three emitted quantum states
to be linearly dependent and, therefore, lie inside a two-dimensional
Hilbert subspace), it does not support losses, and it does not support
decoy states~\cite{decoy03,decoy05a,decoy05b}.
We believe that the analysis of losses and decoy states will work
within our framework (see, e.g.,~\cite{TL17,decoy_sec07,decoy_sec14}),
but we leave their rigorous and precise analysis for future research.
We also believe that our analysis can prove security for practical implementations
of MDI QKD using the reduction techniques introduced by~\cite{mdi_qkd1,mdi_qkd3},
but we leave a detailed analysis of this direction for future research.

In Section~\ref{sec_step}, we explain how the security of practical
implementations of QKD should be rigorously analysed and proved.
In Section~\ref{sec_def}, we fully define the analysed QKD protocol,
and in Section~\ref{sec_proof}, we prove its security;
our final security result (the key rate) is presented
as Corollary~\ref{corollary_security_pm}.
In Section~\ref{sec_rest}, we explain why four source states are likely
too many (in the qubit and two-basis regime) and why we must restrict our protocol
to three states.

\section{\label{sec_step}Step-by-Step Analysis of Practical Implementations of QKD}
Nowadays, despite the enormous progress made on practical security analysis,
a comprehensive method for proving security is still lacking. In many descriptions
of practical implementations of QKD,
while the theoretical model suggested for analysis is close to the practical
implementation, it is naturally not identical, and the reduction from
the practical implementation to the theoretical protocol sometimes uses
hand wavy arguments instead of fully rigorous mathematical modelling and analysis.

For rigorously proving the security of a practical implementation
in the case of a measurement-device-independent (MDI) protocol,
we suggest the following way:
\begin{enumerate}
\item The implementation should be evaluated and tested. In particular,
the emitted quantum states must be repeatedly measured in all aspects,
including determining the modes---frequency spectrum (which includes wavelengths and their relative phases),
polarization, timing and location of emission,
direction of propagation (wave vector), and their degrees of mutual coherence---and performing a full tomography for each mode, thereby
discovering the resulting quantum states. Each resulting quantum
state (for each basis choice and data choice) must be reconstructed
and explicitly written;
this reconstruction is essential for the security analysis.
\item The quantum states must be given as inputs to the security proof.
The security proof then gives us a key rate formula and security parameters.
\item The key rate and security parameters can now be compared to the security
definition. The result of this comparison decides whether security
of the practical implementation has been proved.
\end{enumerate}

In particular, if this process requires any reduction between
the practical implementation and the theoretical model,
the reduction must be rigorous and precise, it must be included as a part of
the proof, and it must be verified to work against any possible attack.

\section{\label{sec_def}Definition of the QKD Protocol}
The QKD protocol we analyse in this paper is a prepare-and-measure protocol,
which is defined as follows:
\begin{enumerate}
\item \label{pm_step_1} Alice and Bob publicly agree on the parameters
of the protocol:
\begin{itemize}
\item Three normalized quantum states
$\{\ket{\gamma_0}_{\sB_i}, \ket{\gamma_1}_{\sB_i}, \ket{\gamma_+}_{\sB_i}\}$
(identical between all rounds $i$)
that can reside in any arbitrary Hilbert space but must be linearly dependent
(and, therefore, must span a two-dimensional Hilbert subspace).
Specifically, we denote
\begin{equation}
\ket{\gamma_+}_{\sB_i} = a \ket{\gamma_0}_{\sB_i} + b \ket{\gamma_1}_{\sB_i},
\end{equation}
where $a, b \in \mathbb{C}$. We require $|a|^2 + |b|^2 > \frac{1}{2}$
or, equivalently,
$\Re (a^\star b \braket{\gamma_0}{\gamma_1}_{\sB_i}) < \frac{1}{4}$.
\item Bob's generalized measurement operators for each round $i$:
\begin{enumerate}
\item $\{M^{\sZ,t}_{\sB_i}\}_{t \in \{0,1\}}$,
which we name ``measurement in the standard basis''
or ``measurement in the $z$ basis'', and
\item $\{M^{\sX,t}_{\sB_i}\}_{t \in \{0,1\}}$,
which we name ``measurement in the conjugate basis''
or ``measurement in the $x$ basis''.
\end{enumerate}
which are defined similarly to~\cite{TL17}.
We note that Bob's measurement operators can be arbitrary and are not required
to be perfectly implemented or perfectly known. However, they influence
the measurement results and the error rate, which influence the protocol's
success probability and key rate.
\item The number $m$ of all rounds (all quantum states sent by Alice to Bob).
\item The probabilities that Alice chooses each ``preparation basis'':
$p^\sA_z$ represents the probability that Alice prepares either
$\ket{\gamma_0}_{\sB_i}$ or $\ket{\gamma_1}_{\sB_i}$
(each of which she chooses with an equal probability, $\frac{p^\sA_z}{2}$),
and $p^\sA_x$ represents the probability that Alice prepares
$\ket{\gamma_+}_{\sB_i}$. We require $p^\sA_z + p^\sA_x = 1$.
\item The probabilities that Bob chooses to measure in each measurement basis:
$p^\sB_z$ (for choosing to measure in the ``$z$ basis'')
and $p^\sB_x$ (for choosing to measure in the ``$x$ basis''), respectively,
such that $p^\sB_z + p^\sB_x = 1$.
\item The numbers $k_1, k_2, k_3, k_4$ of TEST bits required for each pair
of basis choices of Alice and Bob (Z-Z, Z-X, X-Z, and X-X, respectively,
where the first letter ($Z$ or $X$) represents Alice's basis choice,
and the second letter represents Bob's basis choice)
\emph{and} the number $n_1$ of required INFO bits
corresponding to basis choices of Z-Z.
We require $n_1 + k_1 + k_2 + k_3 + k_4 \le m$.
\item The error rate threshold $\delta$ (maximal allowed noise
in TEST-Z-Z and TEST-X-X~bits).
\item The zero rate threshold $\delta_\mathrm{mismatch}$
(maximal allowed rate of ``$+$'' or ``$0$'' results measured by Bob
in TEST-Z-X and TEST-X-Z bits, respectively).
\item The error correction and privacy amplification parameters
described in~\cite{TL17}, including, in particular, the final key length $\ell$.
\end{itemize}
\item \label{pm_step_bases} Alice randomly chooses
a string $\Phi_\sA \in \{0,1\}^m$ of basis choices:
she chooses each bit independently
to have value $0$ with probability $p^\sA_z$
or value $1$ with probability $p^\sA_x$.

Bob randomly chooses a string $\Phi_\sB \in \{0,1\}^m$ of basis choices:
he chooses each bit independently
to have value $0$ with probability $p^\sB_z$
or value $1$ with probability $p^\sB_x$.

In addition, Alice chooses a uniformly random string $R \in \{0,1\}^m$
of the raw bits she prepares and sends (it is only used for rounds
where Alice's basis choice is $0$).

All strings are kept secret.
\item \label{pm_step_prep} For each round $i \in \{1, 2, \ldots, m\}$ of the
protocol, Alice prepares the state dictated by $(\Phi_\sA)_i$ and $R_i$---namely:

\begin{itemize}
\item Alice prepares $\ket{\gamma_0}_{\sB_i}$ if $(\Phi_\sA)_i = 0$ and $R_i = 0$;
\item Alice prepares $\ket{\gamma_1}_{\sB_i}$ if $(\Phi_\sA)_i = 0$ and $R_i = 1$;
\item Alice prepares $\ket{\gamma_+}_{\sB_i}$ if $(\Phi_\sA)_i = 1$
(independently of $R_i$).
\end{itemize}
Alice sends the prepared state to Bob via the quantum channel.
Bob measures each obtained state in the basis dictated by $(\Phi_\sB)_i$
(the ``$z$ basis'' if $(\Phi_\sB)_i = 0$,
or the ``$x$ basis'' if $(\Phi_\sB)_i = 1$)
and puts the measurement result in the string $U \in \{0, 1\}^m$,
which is kept~secret.
\item Bob publicly sends to Alice his basis choice string $\Phi_\sB$.
\item \label{pm_step_min} Alice verifies that
the set $\Sigma \triangleq \{1, 2, \ldots, m\}$
includes at least $n_1 + k_1$ rounds where
Alice chose $z$ and Bob chose $z$ (named ``Z-Z rounds''),
at least $k_2$ ``Z-X rounds'', at least $k_3$ ``X-Z rounds'',
and at least $k_4$ ``X-X rounds''.
If verified, Alice sets the flag $F^\mathrm{min} = \checkmark$;
otherwise, she sets the flag $F^\mathrm{min} = \emptyset$
and aborts the protocol.
\item \label{pm_step_subsets} Alice randomly chooses four subsets
$\Pi_1, \Pi_2, \Pi_3, \Pi_4 \subseteq \Sigma$ of test rounds:
\begin{itemize}
\item $|\Pi_1| = k_1$ is randomly chosen out of all ``Z-Z rounds'' in $\Sigma$,
and it consists of $k_1$ rounds we define as the ``TEST-Z-Z rounds'';
\item $|\Pi_2| = k_2$ is randomly chosen out of all ``Z-X rounds'' in $\Sigma$,
and it consists of $k_2$ rounds we define as the ``TEST-Z-X rounds'';
\item $|\Pi_3| = k_3$ is randomly chosen out of all ``X-Z rounds'' in $\Sigma$,
and it consists of $k_3$ rounds we define as the ``TEST-X-Z rounds'';
\item $|\Pi_4| = k_4$ is randomly chosen out of all ``X-X rounds'' in $\Sigma$,
and it consists of $k_4$ rounds we define as the ``TEST-X-X rounds'',
\end{itemize}
and one subset $\Sigma_1 \subseteq \Sigma$ of information rounds:
\begin{itemize}
\item $|\Sigma_1| = n_1$ is randomly chosen
out of all ``Z-Z rounds'' in $\Sigma \setminus \Pi_1$,
and it consists of $n_1$ rounds we define as the ``INFO rounds''.
\end{itemize}
She publicly sends the five disjoint sets $\Pi_1, \Pi_2, \Pi_3, \Pi_4, \Sigma_1$
to Bob.
\item \label{pm_step_substrings} Each one of Alice and Bob produces five substrings
of their respective bit strings $R,U$:
\begin{itemize}
\item $V^1$ and $W^1$ are the substrings corresponding to $\Pi_1$
(the TEST-Z-Z rounds) of Alice and Bob, respectively;
\item $V^2$ and $W^2$ are the substrings corresponding to $\Pi_2$
(the TEST-Z-X rounds) of Alice and Bob, respectively;
\item $V^3$ and $W^3$ are the substrings corresponding to $\Pi_3$
(the TEST-X-Z rounds) of Alice and Bob, respectively;
\item $V^4$ and $W^4$ are the substrings corresponding to $\Pi_4$
(the TEST-X-X rounds) of Alice and Bob, respectively;
\item $X^1$ and $Y^1$ are the substrings corresponding to $\Sigma_1$
(the INFO rounds) of Alice and Bob, respectively.
\end{itemize}
\item \label{pm_step_pe} Alice sends $V^1, V^4$ to Bob, and Bob compares
them to his $W^1, W^4$ and computes the error rates.
If the error rate in either the TEST-Z-Z rounds
\emph{or} the TEST-X-X rounds exceeds $\delta$,
Bob sets $F^\mathrm{pe} = \emptyset$ and aborts the protocol.

In addition, Bob evaluates his bit strings $W^2, W^3$ and computes their zero rates
(namely, the percentages of his ``$+$'' or ``$0$'' measurement results,
respectively). If the zero rate in either the TEST-Z-X rounds
\emph{or} the TEST-X-Z rounds exceeds $\delta_\mathrm{mismatch}$,
Bob sets $F^\mathrm{pe} = \emptyset$ and aborts the protocol.

If both tests pass, Bob sets $F^\mathrm{pe} = \checkmark$,
and the protocol proceeds.
\item \label{pm_step_ec_pa} Alice and Bob perform error correction
and privacy amplification to their secret INFO bits $X^1, Y^1$
in the standard way for BB84 protocols
(described, e.g., in~\cite{TL17}) to obtain their final secret keys.
We note that Alice and Bob generate another flag, $F^\mathrm{ec}$,
and they abort the protocol if $F^\mathrm{ec} = \emptyset$
(see details in~\cite{TL17});
however, if $F^\mathrm{ec} = \checkmark$, the protocol succeeds,
and Alice's and Bob's final secret keys are denoted by
$K_\sA, K_\sB \in \{0, 1\}^\ell$, respectively.
\end{enumerate}

We point out that this is a very general protocol in the lossless qubit regime
because Alice's emitted states
$\{\ket{\gamma_0}_{\sB_i}, \ket{\gamma_1}_{\sB_i}, \ket{\gamma_+}_{\sB_i}\}$
can be \emph{any} states (assuming they are linearly dependent
and satisfy $|a|^2 + |b|^2 > \frac{1}{2}$), even if they lie
inside a very general Hilbert space (which may be infinite-dimensional
or even continuous). Thus, for this security proof to apply, Alice and Bob must
first test their devices, perform a full quantum tomography of their emitted
states, and input the resulting states
$\ket{\gamma_0}_{\sB_i}, \ket{\gamma_1}_{\sB_i}, \ket{\gamma_+}_{\sB_i}$
to the security proof, as described in Section~\ref{sec_step}.

Using pure states $\ket{\gamma_0}_{\sB_i}, \ket{\gamma_1}_{\sB_i},
\ket{\gamma_+}_{\sB_i}$ does not hurt generality
because if Alice sends a mixed state, we can always assume that she also
sends the purifying system (which Eve intercepts and uses): this assumption
is only beneficial to Eve, so it makes our security proof stronger.

\section{\label{sec_proof}Security Proof}
Our security proof is a generalized version of the rigorous, mostly self-contained
security proof presented by~\cite{TL17} for BB84-based protocols.
That security proof uses entropic uncertainty relations to derive
a key rate formula in the finite-key regime, showing a reduction
from the prepare-and-measure protocol to an entanglement-based protocol.
Here, we generalize this approach to apply to our practical protocol
(in the qubit regime) described in Section~\ref{sec_def}.

\subsection{\label{subsec_eb}Equivalent Modified Entanglement-Based Protocol}
We begin our security proof by performing a reduction
to the following modified entanglement-based protocol.
We point out that this protocol does not strictly adhere to standard definitions
of ``entanglement-based'' protocols because it requires Alice to prepare
a specific entangled state, measure some portions of it, and send other portions
to Bob (which Eve can attack). Therefore, it is similar to prepare-and-measure
protocols. Nevertheless, this protocol is entanglement-based
in the narrowest sense because it allows Alice
to delay her measurements (on some portions of her state)
and relies on the resulting entanglement for proving security.

Therefore, we call it a ``modified entanglement-based protocol'',
and it is defined as~follows:
\begin{enumerate}
\item \label{eb_step_1} Alice and Bob publicly agree on the parameters
of the protocol:
\begin{itemize}
\item Three normalized quantum states
$\{\ket{\gamma_0}_{\sB_i}, \ket{\gamma_1}_{\sB_i}, \ket{\gamma_+}_{\sB_i}\}$
(identical between all rounds $i$)
that can reside in any arbitrary Hilbert space but must be linearly dependent
(and, therefore, must span a two-dimensional Hilbert subspace).
Specifically, we denote
\begin{equation}\label{eq:gamma_+}
\ket{\gamma_+}_{\sB_i} = a \ket{\gamma_0}_{\sB_i} + b \ket{\gamma_1}_{\sB_i},
\end{equation}
where $a, b \in \mathbb{C}$. We require $|a|^2 + |b|^2 > \frac{1}{2}$ or,
equivalently, $\Re (a^\star b \braket{\gamma_0}{\gamma_1}_{\sB_i}) < \frac{1}{4}$.
\\
We also denote the following parameter $T$:
\begin{equation}
T \triangleq |a|^2 + |b|^2\label{eq:T}
\end{equation}
(so $T > \frac{1}{2}$, or $2T - 1 > 0$),
and a resulting \emph{fourth} quantum state $\ket{\gamma_-}_{\sB_i}$:
\begin{equation}
\ket{\gamma_-}_{\sB_i} \triangleq \frac{b^\star \ket{\gamma_0}_{\sB_i}
- a^\star \ket{\gamma_1}_{\sB_i}}{\sqrt{2T - 1}}.\label{eq:gamma_-}
\end{equation}
\begin{Lemma}
If $\ket{\gamma_0}_{\sB_i}, \ket{\gamma_1}_{\sB_i}, \ket{\gamma_+}_{\sB_i}$
are all normalized, then $\ket{\gamma_-}_{\sB_i}$ is normalized, too.
\end{Lemma}
\begin{proof}
\begin{eqnarray}
1 = \braket{\gamma_+}{\gamma_+}_{\sB_i}
&=& |a|^2 \braket{\gamma_0}{\gamma_0}_{\sB_i}
+ |b|^2 \braket{\gamma_1}{\gamma_1}_{\sB_i}
+ 2 \Re (a^\star b \braket{\gamma_0}{\gamma_1}_{\sB_i}) \nonumber \\
&=& T + 2 \Re (a^\star b \braket{\gamma_0}{\gamma_1}_{\sB_i}), \\
\braket{\gamma_-}{\gamma_-}_{\sB_i}
&=& \frac{|b|^2 \braket{\gamma_0}{\gamma_0}_{\sB_i}
+ |a|^2 \braket{\gamma_1}{\gamma_1}_{\sB_i}
- 2 \Re (b a^\star \braket{\gamma_0}{\gamma_1}_{\sB_i})}{2T - 1} \nonumber \\
&=& \frac{T - 2 \Re (a^\star b \braket{\gamma_0}{\gamma_1}_{\sB_i})}{2T - 1}
= \frac{T - (1 - T)}{2T - 1} = \frac{2T - 1}{2T - 1} = 1 
\end{eqnarray}
\end{proof}
\item Inside a separate qubit space
$\mathcal{H}_{\sA_i} \triangleq \Span\{\ket{0}_{\sA_i}, \ket{1}_{\sA_i}\}$,
two orthonormal quantum states (using the same $a,b \in \mathbb{C}$
and $T$ as above):
\begin{eqnarray}
\ket{\xi_+}_{\sA_i} &\triangleq& \frac{a^\star \ket{0}_{\sA_i}
+ b^\star \ket{1}_{\sA_i}}{\sqrt{T}},\label{eq:xi_+} \\
\ket{\xi_-}_{\sA_i} &\triangleq& \frac{b \ket{0}_{\sA_i}
- a \ket{1}_{\sA_i}}{\sqrt{T}},\label{eq:xi_-}
\end{eqnarray}
leading to two orthonormal measurement bases
(representing standard, projective quantum measurements) of Alice
for each round $i$:
\begin{enumerate}
\item $\{\ket{0}_{\sA_i}, \ket{1}_{\sA_i}\}$,
which we name ``the standard basis'' or ``the $z$ basis'', and
\item $\{\ket{\xi_+}_{\sA_i}, \ket{\xi_-}_{\sA_i}\}$,
which we name ``the conjugate basis'' or ``the $x$ basis''.
\end{enumerate}
\item Bob's generalized measurement operators for each round $i$:
\begin{enumerate}
\item $\{M^{\sZ,t}_{\sB_i}\}_{t \in \{0,1\}}$,
which we name ``measurement in the standard basis''
or ``measurement in the $z$ basis'', and
\item $\{M^{\sX,t}_{\sB_i}\}_{t \in \{0,1\}}$,
which we name ``measurement in the conjugate basis''
or ``measurement in the $x$ basis''.
\end{enumerate}
which are defined similarly to~\cite{TL17}.
We note that Bob's measurement operators can be arbitrary and are not required
to be perfectly implemented or perfectly known. However, they influence
the measurement results and the error rate, which influence the protocol's
success probability and key rate.
\item The number $M'$ of all rounds (all quantum states sent by Alice to Bob).
\item The required number $m$ of rounds where Alice does not tell Bob to discard
(see Step~\ref{eb_step_prep}).
\item The probabilities that Alice chooses to measure in each measurement basis:
$p'^\sA_z$ (for choosing to measure in the ``$z$ basis'')
and $p'^\sA_x$ (for choosing to measure in the ``$x$ basis''), respectively,
such that $p'^\sA_z + p'^\sA_x = 1$.
\item The probabilities that Bob chooses to measure in each measurement basis:
$p'^\sB_z$ (for choosing to measure in the ``$z$ basis'')
and $p'^\sB_x$ (for choosing to measure in the ``$x$ basis''), respectively,
such that $p'^\sB_z + p'^\sB_x = 1$.
\item The numbers $k_1, k_2, k_3, k_4$ of TEST bits required for each pair
of basis choices of Alice and Bob (Z-Z, Z-X, X-Z, and X-X, respectively,
where the first letter ($Z$ or $X$) represents Alice's basis choice,
and the second letter represents Bob's basis choice)
\emph{and} the number $n_1$ of required INFO bits
corresponding to basis choices of Z-Z.
We require $n_1 + k_1 + k_2 + k_3 + k_4 \le m$.
\item The error rate threshold $\delta$ (maximal allowed noise
in TEST-Z-Z and TEST-X-X~bits).
\item The zero rate threshold $\delta_\mathrm{mismatch}$
(maximal allowed rate of ``$+$'' or ``$0$'' results measured by Bob
in TEST-Z-X and TEST-X-Z bits, respectively).
\item The error correction and privacy amplification parameters
described in~\cite{TL17}, including, in particular, the final key length $\ell$.
\end{itemize}
\item \label{eb_step_bases} Alice randomly chooses
a string $\Phi_\sA \in \{0,1\}^{M'}$ of basis choices:
she chooses each bit independently
to have value $0$ with probability $p'^\sA_z$ or
value $1$ with probability $p'^\sA_x$.

Bob randomly chooses a string $\Phi_\sB \in \{0,1\}^{M'}$ of basis choices:
he chooses each bit independently
to have value $0$ with probability $p'^\sB_z$ or
value $1$ with probability $p'^\sB_x$.

Both strings are kept secret.
\item \label{eb_step_prep} For each round $i \in \{1, 2, \ldots, M'\}$
of the protocol, Alice generates the following entangled state:
\begin{equation}
\ket{\Psi}_{\sA_i \sB_i} \triangleq \frac{\ket{0}_{\sA_i} \ket{\gamma_0}_{\sB_i}
+ \ket{1}_{\sA_i} \ket{\gamma_1}_{\sB_i}}{\sqrt{2}}
= \frac{\ket{\xi_+}_{\sA_i} \ket{\gamma_+}_{\sB_i}
+ \sqrt{2T - 1} \ket{\xi_-}_{\sA_i} \ket{\gamma_-}_{\sB_i}}{\sqrt{2T}}
\end{equation}
(this equality between its two representations can be proven algebraically).
In other words, Alice generates the state
\begin{equation}
\otimes_{i = 1}^{M'} \ket{\Psi}_{\sA_i \sB_i}
\end{equation}
consisting of the $M'$ quantum systems $\sA_1, \sA_2, \ldots, \sA_{M'}$
(one system for each round $i$).

For each round $i$, if $(\Phi_\sA)_i = 1$
(namely, if Alice will have to measure this round in the ``$x$ basis''),
Alice measures subsystem $\sA_i$ in the ``$x$ basis''
$\{\ket{\xi_+}_{\sA_i}, \ket{\xi_-}_{\sA_i}\}$.
(If $(\Phi_\sA)_i = 0$, she delays measurement to Step~\ref{eb_step_a_mes}.)
Alice then defines the following bit string $D \in \{0,1\}^{M'}$:
\begin{equation}
D_i \triangleq \begin{cases}
1 & \text{If $(\Phi_\sA)_i = 1$ and Alice measures ``$\xi_-$'' in round $i$} \\
0 & \text{Otherwise, (either $(\Phi_\sA)_i = 0$,
or Alice measures ``$\xi_+$'' in round $i$)} \end{cases}
\end{equation}
Alice publicly sends to Bob the string $D$.
This means that for each round $i$, Alice tells Bob (and Eve)
whether she obtained the measurement result ``$\xi_-$'';
however, if she did not obtain the measurement result ``$\xi_-$'',
she \emph{does not} expose the measurement result (if any) or the chosen basis.

Alice and Bob discard and ignore all rounds where $D_i = 1$,
which we name the ``discarded rounds''.
However, for all the ``non-discarded rounds'' (rounds where $D_i = 0$),
Alice sends to Bob the subsystem $\sB_i$ via the quantum channel,
which can be attacked by Eve.
\item Bob publicly sends to Alice his basis choice string $\Phi_\sB$.
\item Alice denotes the set of rounds that were not discarded by her
as $\Omega \subseteq \{1, 2, \ldots, M'\}$
(namely, $\Omega \triangleq \{1 \le i \le M' \mid D_i = 0\}$).
Alice verifies that at least $m$ rounds appear in $\Omega$,
in which case she sets the flag $F^\mathrm{sift'} = \checkmark$ and
publishes the set $\Sigma \subseteq \Omega$
consisting of the first $m$ rounds appearing in $\Omega$
(which are the first $m$ non-discarded rounds).
Otherwise (if fewer than $m$ rounds appear in $\Omega$),
Alice sets the flag $F^\mathrm{sift'} = \emptyset$
and aborts the protocol.

The two next steps are \emph{completely identical}
to Steps~\ref{pm_step_min} and~\ref{pm_step_subsets}
of the original prepare-and-measure protocol described in Section~\ref{sec_def}:
\item \label{eb_step_min} Alice verifies that $\Sigma$ includes
at least $n_1 + k_1$ rounds where
Alice chose $z$ and Bob chose $z$ (named ``Z-Z rounds''),
at least $k_2$ ``Z-X rounds'', at least $k_3$ ``X-Z rounds'',
and at least $k_4$ ``X-X rounds''.
If verified, Alice sets the flag $F^\mathrm{min} = \checkmark$;
otherwise, she sets the flag $F^\mathrm{min} = \emptyset$
and aborts the protocol.
\item \label{eb_step_subsets} Alice randomly chooses four subsets
$\Pi_1, \Pi_2, \Pi_3, \Pi_4 \subseteq \Sigma$ of test rounds:
\begin{itemize}
\item $|\Pi_1| = k_1$ is randomly chosen out of all ``Z-Z rounds'' in $\Sigma$,
and it consists of $k_1$ rounds we define as the ``TEST-Z-Z rounds'';
\item $|\Pi_2| = k_2$ is randomly chosen out of all ``Z-X rounds'' in $\Sigma$,
and it consists of $k_2$ rounds we define as the ``TEST-Z-X rounds'';
\item $|\Pi_3| = k_3$ is randomly chosen out of all ``X-Z rounds'' in $\Sigma$,
and it consists of $k_3$ rounds we define as the ``TEST-X-Z rounds'';
\item $|\Pi_4| = k_4$ is randomly chosen out of all ``X-X rounds'' in $\Sigma$,
and it consists of $k_4$ rounds we define as the ``TEST-X-X rounds'',
\end{itemize}
and one subset $\Sigma_1 \subseteq \Sigma$ of information rounds:
\begin{itemize}
\item $|\Sigma_1| = n_1$ is randomly chosen
out of all ``Z-Z rounds'' in $\Sigma \setminus \Pi_1$,
and it consists of $n_1$ rounds we define as the ``INFO rounds''.
\end{itemize}
She publicly sends the five disjoint sets $\Pi_1, \Pi_2, \Pi_3, \Pi_4, \Sigma_1$
to Bob.
\item \label{eb_step_a_mes} \textls[-25]{Alice measures all quantum systems $\sA_i$
for which $(\Phi_\sA)_i = 0$ in the ``$z$ basis''}
$\{\ket{0}_{\sA_i}, \ket{1}_{\sA_i}\}$.
She puts all her measurement results (from both this step and
Step~\ref{eb_step_prep}) in the string $R \in \{0, 1\}^{M'}$, which is kept secret.
\item \label{eb_step_b_mes} Bob measures all his non-discarded quantum systems
in the bases dictated by $\Phi_\sB$
(the ``$z$ basis'' if $(\Phi_\sB)_i = 0$,
or the ``$x$ basis'' if $(\Phi_\sB)_i = 1$)
and puts his measurement results in the string $U \in \{0, 1\}^{M'}$,
which is kept secret.

The rest of the protocol is \emph{completely identical} to the last steps
of the original prepare-and-measure protocol described in Section~\ref{sec_def}
(in its Steps~\ref{pm_step_substrings}--\ref{pm_step_ec_pa}):
\item \label{eb_step_substrings} Each one of Alice and Bob produces five substrings
of their respective bit strings $R,U$:
\begin{itemize}
\item $V^1$ and $W^1$ are the substrings corresponding to $\Pi_1$
(the TEST-Z-Z rounds) of Alice and Bob, respectively;
\item $V^2$ and $W^2$ are the substrings corresponding to $\Pi_2$
(the TEST-Z-X rounds) of Alice and Bob, respectively;
\item $V^3$ and $W^3$ are the substrings corresponding to $\Pi_3$
(the TEST-X-Z rounds) of Alice and Bob, respectively;
\item $V^4$ and $W^4$ are the substrings corresponding to $\Pi_4$
(the TEST-X-X rounds) of Alice and Bob, respectively;
\item $X^1$ and $Y^1$ are the substrings corresponding to $\Sigma_1$
(the INFO rounds) of Alice and Bob, respectively.
\end{itemize}
\item \label{eb_step_pe} Alice sends $V^1, V^4$ to Bob, and Bob compares
them to his $W^1, W^4$ and computes the error rates.
If the error rate in either the TEST-Z-Z rounds
\emph{or} the TEST-X-X rounds exceeds $\delta$,
Bob sets $F^\mathrm{pe} = \emptyset$ and aborts the protocol.

In addition, Bob evaluates his bit strings $W^2, W^3$ and computes their zero rates
(namely, the percentages of his ``$+$'' or ``$0$'' measurement results,
respectively). If the zero rate in either the TEST-Z-X rounds
\emph{or} the TEST-X-Z rounds exceeds $\delta_\mathrm{mismatch}$,
Bob sets $F^\mathrm{pe} = \emptyset$ and aborts the protocol.

If both tests pass, Bob sets $F^\mathrm{pe} = \checkmark$,
and the protocol proceeds.
\item \label{eb_step_ec_pa} Alice and Bob perform error correction
and privacy amplification to their secret INFO bits $X^1, Y^1$
in the standard way for BB84 protocols
(described, e.g., in~\cite{TL17}) to obtain their final secret keys.
We note that Alice and Bob generate another flag, $F^\mathrm{ec}$,
and they abort the protocol if $F^\mathrm{ec} = \emptyset$
(see details in~\cite{TL17});
however, if $F^\mathrm{ec} = \checkmark$, the protocol succeeds,
and Alice's and Bob's final secret keys are denoted by
$K_\sA, K_\sB \in \{0, 1\}^\ell$, respectively.
\end{enumerate}

In Subsection~\ref{subsec_security}, we prove security of this protocol,
and in Subsection~\ref{subsec_eb_red}, we prove the reduction
to be correct---namely, we prove that security of the above protocol implies
security of the original protocol.

\subsection{\label{subsec_security}Security Proof for the Modified
Entanglement-Based Protocol}
Our security proof is a generalization of Section~6 of~\cite{TL17}
(which proves security of an entanglement-based version
of BB84~\cite{BB84,ekert91,BBM92}),
requiring a few modifications of their proof.

The proof of~\cite{TL17} is based on an \emph{entropic uncertainty relation}
which, roughly speaking, links two quantities:
the smooth min-entropy of Alice's data conditioned on Eve's data
(denoted $H_\mathrm{min}^\epsilon(\sA|\sE)$)
and the smooth max-entropy of Alice's data conditioned on Bob's data
(denoted $H_\mathrm{max}^\epsilon(\sA|\sB)$).
Generally speaking, these entropies are measures of \emph{uncertainty}:
they capture the number of bits in Alice's system $\sA$
that are unknown to either Eve or Bob, respectively.
Intuitively (and imprecisely),
the smooth min-entropy $H_\mathrm{min}^\epsilon(\sA|\sE)$ describes
the number of secret bits that can be \emph{extracted} from Alice's system $\sA$
and will be completely secret even from Eve
(or from anyone that has Eve's system $\sE$),
while the smooth max-entropy $H_\mathrm{max}^\epsilon(\sA|\sB)$ describes
the number of \emph{extra} information bits that Bob will have to receive
from Alice if he wants to have full information on her system $\sA$
(which is roughly equivalent to asking how much information Alice would need
to send to Bob during the error correction procedure).

Roughly speaking, the entropic uncertainty relation used by~\cite{TL17}
shows a lower bound on
$H_\mathrm{min}^\epsilon(\sA|\sE) + H_\mathrm{max}^\epsilon(\sA|\sB)$.
Then, their proof upper-bounds $H_\mathrm{max}^\epsilon(\sA|\sB)$
by bounding the error rate between Alice and Bob using a law of large numbers,
which shows it is unlikely that the error rate on TEST bits is less than $\delta$
while the error rate on INFO bits
is more than $\delta + \nu$. (Intuitively,
the smoothness parameter $\epsilon$ means that we do not necessarily use
the original quantum state given as an input to the entropy,
but we may use any quantum state up to distance $\epsilon$ from it.
In our case, for example, $\epsilon^2$ represents the maximal probability
that the law of large numbers is violated---namely, the maximal probability
that the true error rate in the INFO bits is much higher
than the error rate observed in the TEST bits.
Using the smooth min- and max-entropy allows us to upper-bound the impact
of this unwanted possibility.)
The combination of these two results implies
a lower bound on $H_\mathrm{min}^\epsilon(\sA|\sE)$,
and this lower bound immediately gives us the protocol's key rate using
the Leftover Hashing Lemma~\cite{renner_thesis08}
(which intuitively says that roughly $H_\mathrm{min}^\epsilon(\sA|\sE)$ bits,
known to Alice and completely secret from Eve, can be extracted from Alice's system
using a standard procedure of privacy~amplification).

We mainly modify the first two parts of~\cite{TL17}'s proof:
the entropic uncertainty relation
and the use of the law of large numbers.
In addition, we need to justify security of our sifting step.

\subsubsection{\label{subsubsec_sifting}{The Sifting Step}}
The sifting step of our modified entanglement-based protocol
does not appear in the entanglement-based protocol
of~\cite{TL17}. Therefore, we must prove that it does not hurt security
by showing that it keeps Alice's and Bob's basis choice strings
$\Phi_\sA, \Phi_\sB$ independent of the other~systems.

Each bit of $\Phi_\sA, \Phi_\sB$ is chosen independently
(with probabilities $p'^\sA_z$, $p'^\sA_x$, $p'^\sB_z$, and $p'^\sB_x$,
respectively). Moreover, the state that Alice generates for each round is as follows:
\begin{equation}
\ket{\Psi}_{\sA_i \sB_i} \triangleq \frac{\ket{0}_{\sA_i} \ket{\gamma_0}_{\sB_i}
+ \ket{1}_{\sA_i} \ket{\gamma_1}_{\sB_i}}{\sqrt{2}}
= \frac{\ket{\xi_+}_{\sA_i} \ket{\gamma_+}_{\sB_i}
+ \sqrt{2T - 1} \ket{\xi_-}_{\sA_i} \ket{\gamma_-}_{\sB_i}}{\sqrt{2T}},
\end{equation}
so we observe that if $(\Phi_\sA)_i = 0$
(namely, if Alice chooses the ``$z$ basis''),
Alice obtains the ``$0$'' and ``$1$'' results
with equal conditional probabilities ($\frac{1}{2}$);
and if $(\Phi_\sA)_i = 1$ (namely, if Alice chooses the ``$x$ basis''),
Alice obtains the ``$\xi_+$'' result with conditional probability $\frac{1}{2T}$
and obtains the ``$\xi_-$'' result with conditional probability $\frac{2T-1}{2T}$.
We conclude the following:
\begin{eqnarray}
\pr_{\sA_i}(0) = \pr_{\sA_i}(1) &=& p'^\sA_z \cdot \frac{1}{2}
= \frac{p'^\sA_z}{2}, \\
\pr_{\sA_i}(\xi_+) &=& p'^\sA_x \cdot \frac{1}{2T} = \frac{p'^\sA_x}{2T}, \\
\pr_{\sA_i}(\xi_-) &=& p'^\sA_x \cdot \frac{2T-1}{2T} = \frac{p'^\sA_x (2T-1)}{2T}.
\end{eqnarray}

To justify the sifting step, we notice that this probabilistic process
can equivalently be described as the following two-stage process:
\begin{enumerate}
\item \label{sifting_stage_1} First, for each round $i$, Alice determines
whether round $i$ is discarded ($D_i = 1$) or not ($D_i = 0$).
The round is discarded if and only if Alice measures ``$\xi_-$''; therefore,
\begingroup\makeatletter\def\f@size{9.5}\check@mathfonts
\def\maketag@@@#1{\hbox{\m@th\fontsize{10}{10}\selectfont\normalfont#1}}
\begin{eqnarray}
\pr_{\sA_i}(D_i = 1) &=& \pr_{\sA_i}(\xi_-) = \frac{p'^\sA_x (2T-1)}{2T}, \\
\pr_{\sA_i}(D_i = 0) &=& \pr_{\sA_i}(0) + \pr_{\sA_i}(1) + \pr_{\sA_i}(\xi_+)
= 2 \cdot \frac{p'^\sA_z}{2} + \frac{p'^\sA_x}{2T}
= p'^\sA_z + \frac{p'^\sA_x}{2T}.
\end{eqnarray}
\endgroup
\textls[-15]{Remember that we define $\Omega \triangleq \{1 \le i \le M' \mid D_i = 0\}$
as the set of non-discarded~rounds.}
\item \label{sifting_stage_2}Then, for each round $i$ in $\Omega$
(each non-discarded round), Alice determines the basis:
\begin{eqnarray}
\pr_{\sA_i}\left[(\Phi_\sA)_i = 0 \mid D_i = 0\right]
&=& \frac{\pr_{\sA_i}(0) + \pr_{\sA_i}(1)}{\pr_{\sA_i}(D_i = 0)}
= \frac{2 \cdot \frac{p'^\sA_z}{2}}{p'^\sA_z + \frac{p'^\sA_x}{2T}}
= \frac{p'^\sA_z}{p'^\sA_z + \frac{p'^\sA_x}{2T}},\label{eq:eb_phiA_0} \\
\pr_{\sA_i}\left[(\Phi_\sA)_i = 1 \mid D_i = 0\right]
&=& \frac{\pr_{\sA_i}(\xi_+)}{\pr_{\sA_i}(D_i = 0)}
= \frac{\frac{p'^\sA_x}{2T}}{p'^\sA_z + \frac{p'^\sA_x}{2T}}.\label{eq:eb_phiA_1}
\end{eqnarray}
These probabilities are \emph{independent} between the rounds:
namely, the basis is determined independently for each non-discarded round.
\end{enumerate}

Note that this equivalence is only correct with respect to the \emph{probability
distribution}; the above process does not describe a physical process,
but a virtual process that cannot be applied in practice
and only gives the same probability distribution over $\Phi_A$ and $R$.
(This point can be counter-intuitive: from a quantum theory's point of view,
the choice of discarded and non-discarded rounds obviously depends
on the basis chosen for measurement, yet from a probabilistic point of view,
the process can be divided to the two above stages and still give us
an identical probability distribution.)

We notice that both the flag $F^\mathrm{sift'}$ (which notes whether there are at
least $m$ rounds in $\Omega$---namely, whether at least $m$ rounds were not
discarded) and the set $\Sigma$ (which represents the first $m$ rounds in $\Omega$)
\emph{only} depend on stage~\ref{sifting_stage_1}:
namely, from the probabilistic point of view,
both the passing of the sifting test ($F^\mathrm{sift'}$)
and the choice of the $m$ relevant rounds ($\Sigma$)
depend only on the choice of discarded rounds in stage~\ref{sifting_stage_1},
and they are both completely independent of the bases $\Phi_\sA$ chosen
for the non-discarded rounds in stage~\ref{sifting_stage_2}.
The bases of the non-discarded rounds are chosen independently
with the predetermined probabilities computed
in Equations~\eqref{eq:eb_phiA_0} and \eqref{eq:eb_phiA_1}.

For this reason, in the rest of the analysis, we can treat $\Phi_\sA$ and $\Phi_\sB$
(more precisely, their restrictions to the $m$ non-discarded rounds in $\Sigma$)
as \emph{completely independent} of Eve's attack.
In other words, Eve's attack is applied independently of Alice's and Bob's
chosen bases (or their actually used bases) because Eve is only given access to
the discarding string $D$ which is completely independent of the bases
in the non-discarded rounds.
This result is crucial for the application of the law of large numbers
to hypothetical protocols in Subsubsection~\ref{subsubsec_lln}.

Furthermore, we notice that our modified entanglement-based protocol
actually acts in the following way regarding the choice of bases
and TEST and INFO bits inside $\Sigma$:
\begin{enumerate}
\item In stage~\ref{sifting_stage_2} (included in
Step~\ref{eb_step_bases} of the protocol), Alice and Bob determine
the bases of the $m$ non-discarded rounds in $\Sigma$,
chosen randomly and independently for each round in $\Sigma$.
\item In Step~\ref{eb_step_min} of the protocol, Alice verifies
that $\Sigma$ has sufficient numbers of rounds corresponding to each pair of bases
(namely, at least $n_1 + k_1$ ``Z-Z rounds'', at least $k_2$ ``Z-X rounds'',
at least $k_3$ ``X-Z rounds'', and at least $k_4$ ``X-X rounds'').
For simplicity, let us denote the ``Z-Z rounds'' by $\Sigma_{\sZ,\sZ}$,
the ``Z-X rounds'' by $\Sigma_{\sZ,\sX}$,
the ``X-Z rounds'' by $\Sigma_{\sX,\sZ}$,
and the ``X-X rounds'' by $\Sigma_{\sX,\sX}$;
here, Alice verifies that $|\Sigma_{\sZ,\sZ}| \ge n_1 + k_1$,
$|\Sigma_{\sZ,\sX}| \ge k_2$,
$|\Sigma_{\sX,\sZ}| \ge k_3$,
and $|\Sigma_{\sX,\sX}| \ge k_4$. We condition on passing this
verification---namely, we evaluate the conditional probabilities on
$F^\mathrm{min} = \checkmark$.
\item In Step~\ref{eb_step_subsets} of the protocol,
Alice uniformly and randomly chooses the corresponding disjoint subsets
$\Sigma_1, \Pi_1 \subseteq \Sigma_{\sZ,\sZ}$,
$\Pi_2 \subseteq \Sigma_{\sZ,\sX}$,
$\Pi_3 \subseteq \Sigma_{\sX,\sZ}$, and $\Pi_4 \subseteq \Sigma_{\sX,\sX}$
(of sizes $n_1$, $k_1$, $k_2$, $k_3$, and $k_4$, respectively).
This effectively discards the other $m - n_1 - k_1 - k_2 - k_3 - k_4$ rounds
in $\Sigma \setminus (\Sigma_1 \cup \Pi_1 \cup \Pi_2 \cup \Pi_3 \cup \Pi_4)$,
because their basis choices and measurement results are completely ignored
by the rest of the protocol.
\end{enumerate}

Combining these three steps and conditioning on $F^\mathrm{min} = \checkmark$
(namely, conditioning on passing the verification of the second step),
this process is equivalent from the probabilistic point of view
to uniformly and randomly choosing five disjoint subsets
$\Sigma_1, \Pi_1, \Pi_2, \Pi_3, \Pi_4 \subseteq \Sigma$
of sizes $n_1$, $k_1$, $k_2$, $k_3$, and $k_4$, respectively
(out of the $m$-sized set $\Sigma$),
and letting this uniform choice dictate the choice of measurement probabilities
(Z-Z, Z-Z, Z-X, X-Z, and X-X, respectively).
This equivalence results from the complete symmetry of the above three steps,
which have no dependence whatsoever
on the identifying number of each round inside $\Sigma$.
This observation, too, is crucial for the application of the law of large numbers
in Subsubsection~\ref{subsubsec_lln}.

\subsubsection{Entropic Uncertainty Relation}
The security proof in~\cite{TL17} uses the uncertainty relation described
in Proposition~4 of~\cite{TL17}, which is equivalent to Corollary~7.4
of~\cite{tomamichel_thesis12}. This uncertainty relation
uses the symmetry of the BB84 protocol with respect to the chosen basis
of the INFO bits: namely, it uses the property that each INFO bit in BB84
is measured in a uniformly random basis, independently of the TEST bits' results
and bases. Since a similar property does not apply to our protocol
(because all our INFO bits are measured in the ``$z$ basis''),
we must use a different uncertainty relation.

Adopting an approach similar to~\cite{TLGR12},
we use Theorem~7.2 of~\cite{tomamichel_thesis12} as
the generalized entropic uncertainty relation
(using slightly different notations compared to~\cite{tomamichel_thesis12}):
\begin{Theorem}\label{theorem_uncertainty}
(Theorem~7.2 of~\cite{tomamichel_thesis12})
Given $\epsilon \ge 0$ and a non-normalized state $\rho_{\sA\sR\sS}$
over the Hilbert space $\mathcal{H}_{\sA\sR\sS}$,
and given two generalized measurement operators
$M_1 = \{M_1^x\}_x ~ , ~ M_2 = \{M_2^{x'}\}_{x'}$
on $\mathcal{H}_\sA$ and a projective measurement $\{P^p\}_p$
on $\mathcal{H}_\sA$, the two post-measurement states
\begin{eqnarray}
\sigma_{\sX\sP\sR} &=& \sum_{x,p} \ket{x}_\sX \bra{x}_\sX \otimes
\ket{p}_\sP \bra{p}_\sP \otimes
\tr_{\sA\sS}\left(M_1^x P^p \rho_{\sA\sR\sS} P^p (M_1^x)^\dagger\right), \\
\sigma'_{\sX'\sP\sS} &=& \sum_{x',p} \ket{x'}_{\sX'} \bra{x'}_{\sX'} \otimes
\ket{p}_\sP \bra{p}_\sP \otimes
\tr_{\sA\sR}\left(M_2^{x'} P^p \rho_{\sA\sR\sS} P^p (M_2^{x'})^\dagger\right)
\end{eqnarray}
satisfy the following inequality (uncertainty relation):
\begin{equation}
H_\mathrm{min}^\epsilon(\sX|\sP\sR)_\sigma
+ H_\mathrm{max}^\epsilon(\sX'|\sP\sS)_{\sigma'}
\ge \log_2 \left( \frac{1}{c_\sP} \right),
\end{equation}
where:
\begin{equation}
c_\sP \triangleq \max_{p,x,x'} \left|\left|M_1^x P^p
(M_2^{x'})^\dagger\right|\right|_\infty^2.
\end{equation}
\end{Theorem}
\begin{proof}
Proved in Subsection~7.3.2 of~\cite{tomamichel_thesis12} as Theorem~7.2.
\end{proof}

The next proposition will give us a similar result to Corollary~5 of~\cite{TL17}
(namely, roughly speaking, a lower bound on the sum
$H_\mathrm{min}^\epsilon(\sA|\sE) + H_\mathrm{max}^\epsilon(\sA|\sB)$),
with one important difference:
the first term $H_\mathrm{min}^\epsilon(\sA|\sE)$ will still refer
to the real QKD protocol (the modified entanglement-based protocol
described in Subsection~\ref{subsec_eb}),
but the second term $H_\mathrm{max}^\epsilon(\sA|\sB)$
will now refer to a hypothetical QKD protocol (still entanglement-based)
where both Alice and Bob measure the INFO bits in the conjugate (``$x$'') basis.
Formally:
\begin{Proposition}\label{prop_uncertainty}
For the modified entanglement-based protocol
described in Subsection~\ref{subsec_eb}, for $\epsilon \ge 0$,
the state $\sigma_{\sX^1 \sY^1 \sV \sW \Pi \Phi_\sA \Phi_\sB
\sF^\mathrm{sift'} \sF^\mathrm{min} \sF^\mathrm{pe} \sE}$ held by Alice, Bob,
and Eve after Step~\ref{eb_step_pe} of the real protocol, and
the state $\sigma'_{\sX'^1 \sY'^1 \sV \sW \Pi \Phi_\sA \Phi_\sB
\sF^\mathrm{sift'} \sF^\mathrm{min} \sF^\mathrm{pe} \sE}$
held by Alice, Bob, and Eve after Step~\ref{eb_step_pe}
of the hypothetical protocol defined below, it holds that
\begin{eqnarray}
&&H_\mathrm{min}^\epsilon(\sX^1 \wedge \sF^\mathrm{pe} = \checkmark|\sV \sW
\Pi \Phi_\sA \Phi_\sB \sE ~ , ~ \sF^\mathrm{sift'} = \sF^\mathrm{min}
= \checkmark)_\sigma \nonumber \\
&+& H_\mathrm{max}^\epsilon(\sX'^1 \wedge \sF^\mathrm{pe} = \checkmark|\sY'^1
~ , ~ \sF^\mathrm{sift'} = \sF^\mathrm{min} = \checkmark)_{\sigma'}
\ge n_1 \log_2 \left( \frac{1}{c} \right),
\end{eqnarray}
where we define $\sV \triangleq (V^1, V^2, V^3, V^4)$,
$\sW \triangleq (W^1, W^2, W^3, W^4)$,
$\Pi \triangleq (\Pi_1, \Pi_2, \Pi_3, \Pi_4, \Sigma_1, D, \Sigma)$,
and $c \triangleq \max\left(|\braket{0}{\xi_+}_{\sA_i}|^2,
|\braket{0}{\xi_-}_{\sA_i}|^2,
|\braket{1}{\xi_+}_{\sA_i}|^2, |\braket{1}{\xi_-}_{\sA_i}|^2\right)$,
using the protocol's notations;
subsystem $\sE$ represents Eve's ancilla (including her quantum data);
and $\sX'^1,\sY'^1$ are the substrings corresponding to the $n_1$ INFO rounds
(namely, the Z-Z rounds in $\Sigma_1$)
\emph{in the hypothetical protocol where both Alice and Bob measure
the INFO bits in the conjugate (``$x$'') basis}.

(The hypothetical protocol only changes the actual measurements performed
by Alice and Bob in Steps~\ref{eb_step_a_mes} and~\ref{eb_step_b_mes}
of the modified entanglement-based protocol.
It does not change any other part of the protocol: in particular,
Alice neither discards INFO rounds where she measured ``$\xi_-$''
in the hypothetical protocol nor notifies Bob about them.)
\end{Proposition}

\begin{proof}
This proof combines modified versions of
the proofs of Corollary~7.4 in Subsection~7.4.2 of~\cite{tomamichel_thesis12}
and Corollary~5 in Subsection~6.2 of~\cite{TL17}.

We choose the measurement operators
$M_1 = \left\{\ket{j}_\sA \bra{j}_\sA \mid j \in \{0, 1\}^{n_1}\right\}$
(i.e., the \linebreak tensor product of $n_1$ copies of the ``$z$ basis'') and
$M_2 = \left\{\ket{j}_\sA \bra{j}_\sA \mid j \in \{\xi_+, \xi_-\}^{n_1}\right\}$
\linebreak (i.e., the tensor product of $n_1$ copies of the ``$x$ basis'')
and the projective\linebreak  measurement  $\{P^p\}_p = \{\ket{\Pi\Phi_\sA\Phi_\sB}
\bra{\Pi\Phi_\sA\Phi_\sB}\}_{\Pi\Phi_\sA\Phi_\sB}$.
It is easy to verify (see, e.g.,~\cite{tomamichel_thesis12}) \linebreak
that $c_\sP$ of Theorem~\ref{theorem_uncertainty} is equal to
$c^{n_1}$, where $c$ was defined in our proposition \linebreak
($c \triangleq \max\left(|\braket{0}{\xi_+}_{\sA_i}|^2,
|\braket{0}{\xi_-}_{\sA_i}|^2,
|\braket{1}{\xi_+}_{\sA_i}|^2, |\braket{1}{\xi_-}_{\sA_i}|^2\right)$).

Then, we apply Theorem~\ref{theorem_uncertainty} to the state
\emph{after} Alice's and Bob's measurements of all TEST bits,
\emph{before} Alice's and Bob's measurements
of the INFO bits (here, it is important that our modified entanglement-based
protocol delays to its Step~\ref{eb_step_a_mes} all Alice's measurements
in the ``$z$ basis'', which include all measurements of the INFO bits),
\emph{conditioned} on the first two tests passing
($\sF^\mathrm{sift'} = \sF^\mathrm{min} = \checkmark$),
and \emph{requiring} the third test to pass ($\sF^\mathrm{pe} = \checkmark$):
(the difference between ``conditioning'' and ``requiring'' in this context
is analogous to the difference between a ``conditional probability'' and
a ``joint probability'', respectively; see~\cite{TL17} for the precise definitions)
\begin{equation}
\rho_{\sA \sB \sV \sW \Pi \Phi_\sA \Phi_\sB
\sF^\mathrm{sift'} \sF^\mathrm{min} \sF^\mathrm{pe} \sE
~ \wedge ~ \sF^\mathrm{pe} = \checkmark
~ \mid ~ \sF^\mathrm{sift'} = \sF^\mathrm{min} = \checkmark},
\end{equation}
and by choosing the systems $\sP = \Pi\Phi_\sA\Phi_\sB ~ , ~
\sR = \sV \sW \sE ~, ~ \sS = \sB$ for Theorem~\ref{theorem_uncertainty}, we obtain the following:
\begin{eqnarray}
&&H_\mathrm{min}^\epsilon(\sX^1 \wedge \sF^\mathrm{pe} = \checkmark|\sV \sW
\sE \Pi \Phi_\sA \Phi_\sB ~ , ~ \sF^\mathrm{sift'} = \sF^\mathrm{min}
= \checkmark)_\sigma \nonumber \\
&+& H_\mathrm{max}^\epsilon(\sX'^1 \wedge \sF^\mathrm{pe} = \checkmark|\sB
\Pi \Phi_\sA \Phi_\sB ~ , ~ \sF^\mathrm{sift'} = \sF^\mathrm{min}
= \checkmark)_{\sigma'} \ge n_1 \log_2 \left( \frac{1}{c} \right),
\end{eqnarray}
where $X^1$ and $X'^1$ represent the measurement results of Alice's INFO bits
in the ``$z$ basis'' and in the ``$x$ basis'', respectively.

Performing a measurement of Bob's INFO bits in subsystem $\sB$
in the ``$x$ basis'' (yielding the bit string $\sY'^1$)
and discarding the classical information systems,
$\Pi\Phi_\sA\Phi_\sB$ are two quantum operations (CPTP) operated exclusively on
subsystems $\sB\Pi\Phi_\sA\Phi_\sB$. According to
the data processing inequality (see, e.g.,~\cite{renner_thesis08,TL17}),
such operations can only increase the~max-entropy:
\begingroup\makeatletter\def\f@size{8}\check@mathfonts
\def\maketag@@@#1{\hbox{\m@th\fontsize{10}{10}\selectfont\normalfont#1}}
\begin{equation}
H_\mathrm{max}^\epsilon(\sX'^1 \wedge \sF^\mathrm{pe} = \checkmark|\sB
\Pi \Phi_\sA \Phi_\sB ~ , ~ \sF^\mathrm{sift'} = \sF^\mathrm{min}
= \checkmark)_{\sigma'} \le
H_\mathrm{max}^\epsilon(\sX'^1 \wedge \sF^\mathrm{pe} = \checkmark|\sY'^1
~ , ~ \sF^\mathrm{sift'} = \sF^\mathrm{min} = \checkmark)_{\sigma'},
\end{equation}
\endgroup
which gives us the desired result:
\begin{eqnarray}
&&H_\mathrm{min}^\epsilon(\sX^1 \wedge \sF^\mathrm{pe} = \checkmark|\sV \sW
\sE \Pi \Phi_\sA \Phi_\sB ~ , ~ \sF^\mathrm{sift'} = \sF^\mathrm{min}
= \checkmark)_\sigma \nonumber \\
&+& H_\mathrm{max}^\epsilon(\sX'^1 \wedge \sF^\mathrm{pe} = \checkmark|\sY'^1
~ , ~ \sF^\mathrm{sift'} = \sF^\mathrm{min} = \checkmark)_{\sigma'}
\ge n_1 \log_2 \left( \frac{1}{c} \right).
\end{eqnarray}
\end{proof}

\subsubsection{\label{subsubsec_lln}The Law of Large Numbers}
Intuitively, the proof of~\cite{TL17} uses a law of large numbers
(Lemma~6 of~\cite{TL17}) to upper-bound the max-entropy
$H_\mathrm{max}^\epsilon(\sX|\sY)$---namely, the max-entropy of Alice's raw key
conditioned on Bob's raw key in the \emph{real} protocol.
However, in our proof, we need
to bound $H_\mathrm{max}^\epsilon(\sX'^1|\sY'^1)$---namely,
the max-entropy of Alice's raw key conditioned on Bob's raw key
in a \emph{hypothetical} protocol where Alice and Bob measure the INFO bits
in the conjugate (``$x$'') basis.
For obtaining this bound, we need to apply the law of large numbers \emph{twice}
to two different hypothetical protocols:
\begin{enumerate}
\item In the \emph{X-X hypothetical protocol} (where both Alice and Bob measure
the INFO bits in the ``$x$ basis''), we can compare the TEST-X-X bits
(where the only non-discarded rounds are those where Alice measured
``$\xi_+$'') to the INFO bits. This way, we can discover the error rate
on the ``$\xi_+$'' bits.
\item For finding the error rate on the ``$\xi_-$'' bits,
we use the \emph{Z-X hypothetical protocol},
where Alice measures the INFO bits in the ``$z$ basis''
while Bob measures them in the ``$x$ basis''.
The following intuitive formula will give us the needed bound:\begingroup\makeatletter\def\f@size{8.2}\check@mathfonts
\def\maketag@@@#1{\hbox{\m@th\fontsize{10}{10}\selectfont\normalfont#1}}
\begin{equation}
\pr(\sA_i = - ~ , ~ \sB_i = +) ~ = ~ \pr(\sA_i = 0 ~ , ~ \sB_i = +)
+ \pr(\sA_i = 1 ~ , ~ \sB_i = +) - \pr(\sA_i = + ~ , ~ \sB_i = +).
\end{equation}
\endgroup
This formula is intuitively trivial because it follows from the following formula:
\begin{eqnarray}
&&\pr(\sA_i = - ~ , ~ \sB_i = +) + \pr(\sA_i = + ~ , ~ \sB_i = +)
~~ = ~~ \pr(\sB_i = +) \nonumber \\
&=& \pr(\sA_i = 0 ~ , ~ \sB_i = +) + \pr(\sA_i = 1 ~ , ~ \sB_i = +).
\end{eqnarray}
Formally, it follows from the independence of Alice's and Bob's operations,
as elaborated in the ``bounding the fourth probability'' portion of the proof of
Proposition~\ref{prop_lln} below.

This idea can be compared with~\cite{lt_qkd_correlated_sources20}'s analysis
of the ``loss tolerant'' protocol (improving on the usual analysis which involves
matrix computations~\cite{lt_qkd14,lt_qkd_exp15,lt_qkd_finite_key15,lt_qkd_gen19}),
but their analysis is more complicated and has several free parameters.
Here, we present a full and precise analysis, leading to an explicit key rate formula
in the single-qubit regime.
\end{enumerate}

Formally, we use the following law of large numbers (Lemma~6 of~\cite{TL17}):
\begin{Lemma}\label{lemma_lln}
(Lemma~6 of~\cite{TL17})
Given a set of $N$ random variables $Z = (Z_1, Z_2, \ldots, Z_N)$, where
each $Z_i$ takes values in $\{0, 1\}$ and $N = a + b$,
and given an \emph{independent}, a uniformly distributed subset
$\Pi \subseteq \{1, 2, \ldots, N\}$ of size $a$, it holds that
\begin{equation}
\pr \left[ \sum_{i \in \Pi} Z_i \le a \delta ~ \wedge ~
\sum_{i \in \overline{\Pi}} Z_i \ge b \cdot (\delta + \nu) \right]
\le e^{-\frac{2 b a^2 \nu^2}{(a + b)(a + 1)}}.
\end{equation}
\end{Lemma}
\begin{proof}
Proved in Subsection~6.3 of~\cite{TL17} as Lemma~6.
\end{proof}

We also use another law of large numbers,
proved in Section~2 of~\cite{Ho63} as Theorem~1:
\begin{Lemma}\label{lemma_lln_indep}
Let $Z_1, \ldots, Z_N$ be independent random variables
with finite first and second moments,
such that $0 \le Z_i \le 1$ for all $1 \le i \le N$.
If $\overline{Z} \triangleq \frac{Z_1 + ... + Z_N}{N}$ is their average and
$\mu \triangleq E[\overline{Z}]$ is the expected value
of $\overline{Z}$, then for any $\nu > 0$,
\begin{equation}
\pr\left[ \overline{Z} - \mu \ge \nu \right] \le e^{-2N\nu^2}.
\end{equation}
\end{Lemma}

Using these Lemmas, we prove the following
(a modified version of Proposition~8 of~\cite{TL17}):
\begin{Proposition}\label{prop_lln}
For the modified entanglement-based protocol
described in Subsection~\ref{subsec_eb},
for the state $\sigma'_{\sX'^1 \sY'^1 \sV \sW \Pi \Phi_\sA \Phi_\sB
\sF^\mathrm{sift'} \sF^\mathrm{min} \sF^\mathrm{pe} \sE}$
defined in Proposition~\ref{prop_uncertainty},
and for error rate threshold $\delta$
and zero rate threshold $\delta_\mathrm{mismatch}$,
if we define for any $0 < \nu \le \frac{1}{2} - \delta$:
\begin{eqnarray}
\delta'(\nu) &\triangleq& \delta_\mathrm{mismatch} + \nu
- \left(\frac{1}{2T} - \nu\right) \cdot (1 - 2 \delta - 2 \nu), \\
\epsilon(\nu) &\triangleq& \sqrt{e^{-2 n_1 \nu^2}
+ e^{-\frac{2 n_1 \cdot \left( \frac{1}{2T} - \nu \right) k_4^2
\nu^2}{\left(k_4 + n_1 \cdot \left( \frac{1}{2T} - \nu \right)\right)(k_4 + 1)}}
+ e^{-\frac{2 n_1 k_2^2 \nu^2}{(k_2 + n_1)(k_2 + 1)}}},
\end{eqnarray}
\textls[-45]{then, for any $0 < \nu \le \frac{1}{2} - \delta$ satisfying
$0 < \delta'(\nu) \le \frac{1}{2}$ and $\epsilon(\nu)^2 < \pr \left[ F^\mathrm{pe}
= \checkmark \mid F^\mathrm{sift'} = F^\mathrm{min} = \checkmark \right]$,}
it holds that
\begin{equation}
H_\mathrm{max}^{\epsilon(\nu)}(\sX'^1 \wedge \sF^\mathrm{pe} = \checkmark \mid
\sY'^1 ~ , ~ \sF^\mathrm{sift'} = \sF^\mathrm{min} = \checkmark)_{\sigma'}
\le n_1 \cdot h_2(\delta'(\nu)),
\end{equation}
where $h_2(x) \triangleq -x \log_2(x) -(1-x) \log_2(1-x)$.
\end{Proposition}
\begin{proof}
Let us define the following event:
\begin{equation}
\Omega_0 \triangleq 1 \left\{ \sum_{i = 1}^{n_1} 1\{ X'^1_i \ne Y'^1_i \} >
n_1 \delta'(\nu) \right\}.
\end{equation}
We need to prove the following probability to be exponentially small:
\begin{eqnarray}
&&\pr \left[ F^\mathrm{pe} = \checkmark ~ \wedge ~ \Omega_0 \mid
F^\mathrm{sift'} = F^\mathrm{min} = \checkmark \right] \nonumber \\
&=& \pr \left[ \sum_{i = 1}^{k_1} 1\{ V^1_i \ne W^1_i \} \le k_1 \delta
~~ \wedge ~~ \sum_{i = 1}^{k_2} 1\{ W^2_i = 0 \} \le k_2 \delta_\mathrm{mismatch}
\right. \nonumber \\
&\wedge& \left.
\sum_{i = 1}^{k_3} 1\{ W^3_i = 0 \} \le k_3 \delta_\mathrm{mismatch} ~~ \wedge ~~
\sum_{i = 1}^{k_4} 1\{ V^4_i \ne W^4_i \} \le k_4 \delta \right. \nonumber \\
&\wedge& \left.
\sum_{i = 1}^{n_1} 1\{ X'^1_i \ne Y'^1_i \} > n_1 \delta'(\nu) ~ \mid ~
F^\mathrm{sift'} = F^\mathrm{min} = \checkmark \right] \nonumber \\
&\le& \pr \left[ \sum_{i = 1}^{k_2} 1\{ W^2_i = 0 \} \le
k_2 \delta_\mathrm{mismatch} ~~ \wedge ~~
\sum_{i = 1}^{k_4} 1\{ V^4_i \ne W^4_i \} \le k_4 \delta \right. \nonumber \\
&\wedge& \left.
\sum_{i = 1}^{n_1} 1\{ X'^1_i \ne Y'^1_i \} > n_1 \delta'(\nu) ~ \mid ~
F^\mathrm{sift'} = F^\mathrm{min} = \checkmark \right].
\end{eqnarray}
Let us remember that $V^2$ and $W^2$ are Alice's and Bob's substrings
corresponding to $\Pi_2$ (the $k_2$ TEST-Z-X rounds);
$V^4$ and $W^4$ are Alice's and Bob's substrings
corresponding to $\Pi_4$ (the $k_4$ TEST-X-X rounds);
and $X'^1$ and $Y'^1$ are Alice's and Bob's substrings
corresponding to $\Sigma_1$ (the $n_1$ INFO rounds)
\emph{in the X-X hypothetical protocol}---namely,
\emph{assuming that both Alice and Bob measured the INFO bits
in the ``$x$ basis''} in Steps~\ref{eb_step_a_mes} and~\ref{eb_step_b_mes}
of the protocol, respectively.

Let us also denote Alice's ``$\xi_+$ rate'' (the percentage of INFO bits which
Alice measures as ``$\xi_+$'') in the X-X hypothetical protocol by $R'_+$---namely,
$R'_+ \triangleq \frac{1}{n_1} \sum_{i = 1}^{n_1} 1\{X'^1_i = 0\}$.
Thus, the probability
$\pr \left[ F^\mathrm{pe} = \checkmark ~ \wedge ~ \Omega_0 \mid
F^\mathrm{sift'} = F^\mathrm{min} = \checkmark \right]$
is bounded by the sum of four probabilities:
\vspace{-12pt}
\begin{adjustwidth}{-\extralength}{0cm}
\begin{eqnarray}
&&\pr \left[ F^\mathrm{pe} = \checkmark ~ \wedge ~ \Omega_0 \mid
F^\mathrm{sift'} = F^\mathrm{min} = \checkmark \right] \nonumber \\
&\le& \pr \left[ R'_+ \le \frac{1}{2T} - \nu ~ \mid ~
F^\mathrm{sift'} = F^\mathrm{min} = \checkmark \right] \nonumber \\
&+& \pr \left[ R'_+ \ge \frac{1}{2T} - \nu
~~ \wedge ~~ \sum_{i = 1}^{k_4} 1\{V_i^4 \ne W_i^4\} \le k_4 \delta
\right. \nonumber \\
&\wedge& \left. \sum_{i = 1}^{n_1} 1\{X'^1_i = 0 ~ \wedge ~ Y'^1_i = 1\} \ge
n_1 R'_+ \cdot (\delta + \nu) ~ \mid ~
F^\mathrm{sift'} = F^\mathrm{min} = \checkmark \right] \nonumber \\
&+& \pr \left[ \sum_{i = 1}^{k_2} 1\{W_i^2 = 0\} \le k_2 \delta_\mathrm{mismatch}
~~ \wedge ~~ \sum_{i = 1}^{n_1} 1\{Y'^1_i = 0\} \ge
n_1 \cdot (\delta_\mathrm{mismatch} + \nu) ~ \mid ~
F^\mathrm{sift'} = F^\mathrm{min} = \checkmark \right] \nonumber \\
&+& \pr \left[ R'_+ \ge \frac{1}{2T} - \nu
~~ \wedge ~~ \sum_{i = 1}^{n_1} 1\{X'^1_i = 0 ~ \wedge ~ Y'^1_i = 1\} \le
n_1 R'_+ \cdot (\delta + \nu) \right. \nonumber \\
&\wedge& \left. \sum_{i = 1}^{n_1} 1\{Y'^1_i = 0\} \le
n_1 \cdot (\delta_\mathrm{mismatch} + \nu)
~~ \wedge ~~ \sum_{i = 1}^{n_1} 1\{X'^1_i \ne Y'^1_i\} > n_1 \delta'(\nu) ~ \mid ~
F^\mathrm{sift'} = F^\mathrm{min} = \checkmark \right].
\end{eqnarray}
\end{adjustwidth}

We now bound each of these four probabilities:

\paragraph{\textbf{Bounding the first probability:}}
We need to bound 
\begin{equation}
\pr \left[ R'_+ \le \frac{1}{2T} - \nu ~ \mid ~
F^\mathrm{sift'} = F^\mathrm{min} = \checkmark \right],
\end{equation}
where $R'_+$ is the ``$\xi_+$'' measurement rate of Alice among the INFO bits
in the X-X hypothetical protocol.
We notice that this rate is only dictated by identical quantum actions performed
by Alice: because Alice measures all INFO bits in the ``$x$ basis''
in the hypothetical protocol, her measurement results are obtained independently
for all rounds and her probability of measuring ``$\xi_+$'' is always
$\frac{1}{2T}$. Namely, Alice's measurement results are $n_1$ independent
random variables $\{Z_i\}_{i=1}^{n_1}$ (with all probabilities conditioned on
$F^\mathrm{sift'} = F^\mathrm{min} = \checkmark$) such that for each $i$:
\begin{equation}
\pr(Z_i = 0 \mid F^\mathrm{sift'} = F^\mathrm{min} = \checkmark)
= \frac{1}{2T} ~~~ , ~~~
\pr(Z_i = 1 \mid F^\mathrm{sift'} = F^\mathrm{min} = \checkmark)
= 1 - \frac{1}{2T} ~~~ .
\end{equation}
Therefore, the expected value of each $Z_i$ is $E[Z_i] = 1 - \frac{1}{2T}$.

We can thus apply Lemma~\ref{lemma_lln_indep} (which applies to $N$ independent
random variables) to the random variables $\{Z_i\}_{i=1}^{n_1}$
with parameters $N = n_1$ and $\mu = E[\overline{Z}] = 1 - \frac{1}{2T}$.
We note that $\overline{Z} = 1 - R'_+$. Therefore, we obtain the following result:
\vspace{-12pt}
\begin{adjustwidth}{-\extralength}{0cm}
\begin{eqnarray}
\pr\left[ R'_+ \le \frac{1}{2T} - \nu ~ \mid ~
F^\mathrm{sift'} = F^\mathrm{min} = \checkmark \right]
&=& \pr\left[ R'_+ - \frac{1}{2T} \le - \nu ~ \mid ~
F^\mathrm{sift'} = F^\mathrm{min} = \checkmark \right] \nonumber \\
&=& \pr\left[ \frac{1}{2T} - R'_+ \ge \nu ~ \mid ~
F^\mathrm{sift'} = F^\mathrm{min} = \checkmark \right] \nonumber \\
&=& \pr\left[ \left( 1 - R'_+ \right) - \left( 1 - \frac{1}{2T} \right)
\ge \nu ~ \mid ~ F^\mathrm{sift'} = F^\mathrm{min} = \checkmark \right]\nonumber \\
&=& \pr\left[ \overline{Z} - \mu \ge \nu ~ \mid ~
F^\mathrm{sift'} = F^\mathrm{min} = \checkmark \right]
\le e^{-2 n_1 \nu^2}.
\end{eqnarray}
\end{adjustwidth}

\paragraph{\textbf{Bounding the second probability:}}
We need to bound
\begin{align}
\pr &\left[ R'_+ \ge \frac{1}{2T} - \nu ~~ \wedge ~~
\sum_{i = 1}^{k_4} 1\{V_i^4 \ne W_i^4\} \le k_4 \delta \right. \nonumber \\
&\wedge ~~ \left. \sum_{i = 1}^{n_1} 1\{X'^1_i = 0 ~ \wedge ~ Y'^1_i = 1\} \ge
n_1 R'_+ \cdot (\delta + \nu) ~ \mid ~
F^\mathrm{sift'} = F^\mathrm{min} = \checkmark \right],
\end{align}
where $V^4$ and $W^4$ are Alice's and Bob's substrings
corresponding to $\Pi_4$ (the $k_4$ TEST-X-X rounds);
$X'^1$ and $Y'^1$ are Alice's and Bob's substrings
corresponding to $\Sigma_1$ (the $n_1$ INFO rounds)
in the X-X hypothetical protocol;
and $R'_+$ is the ``$\xi_+$'' measurement rate of Alice among the INFO bits
in the X-X hypothetical protocol.
We notice that the TEST-X-X rounds in $\Pi_4$ consist \emph{only} of rounds
where Alice measured ``$\xi_+$'' (the other rounds are discarded),
so her recorded bit must be $0$;
therefore, the error event $V_i^4 \ne W_i^4$ is actually equivalent
to $V_i^4 = 0 ~ \wedge ~ W_i^4 = 1$, and the probability is actually
\begin{align}
= \pr &\left[ R'_+ \ge \frac{1}{2T} - \nu ~~ \wedge ~~
\sum_{i = 1}^{k_4} 1\{V_i^4 = 0 ~ \wedge ~ W_i^4 = 1\} \le k_4 \delta
\right. \nonumber \\
&\wedge \left. \sum_{i = 1}^{n_1} 1\{X'^1_i = 0 ~ \wedge ~ Y'^1_i = 1\} \ge
n_1 R'_+ \cdot (\delta + \nu) ~ \mid ~
F^\mathrm{sift'} = F^\mathrm{min} = \checkmark \right].
\end{align}
We notice that all rates are evaluated in the X-X hypothetical protocol;
that in all rounds, both Alice and Bob measure in the ``$x$ basis'';
and that in all rounds taken into account, Alice obtains the ``$\xi_+$'' result.
We thus notice that the quantum behaviour of Alice, Bob, and Eve
is \emph{identical} on all these rounds in the X-X hypothetical protocol
(in particular, $D_i = 0$ for all these rounds, and while the timing of
Alice's measurements may differ between the rounds, this timing is meaningless
from the quantum point of view).

Therefore, we can apply Lemma~\ref{lemma_lln} using the following parameters:
the random variables $Z = (Z_1, Z_2, \ldots, Z_N)$
represent the condition that Alice's bit is $0$ and Bob's bit is $1$
(namely, $Z_i$ represents the evaluation of the condition
$V_i^4 = 0 ~ \wedge ~ W_i^4 = 1$
or $X'^1_i = 0 ~ \wedge ~ Y'^1_i = 1$, respectively);
the sampled subset $\Pi$ includes the $a = k_4$
TEST-X-X rounds in the $\Pi_4$ subset chosen by the protocol,
and the rest $\overline{\Pi}$ includes the $b = n_1 R'_+$ INFO rounds
in the $\Sigma_1$ subset chosen by the protocol
where Alice obtains the ``$\xi_+$'' measurement result.
The sampled susbet $\Pi$ is completely independent of Bob's measurement results
(that are dictated solely by Eve's transmitted states and Alice's results
in the ``$x$ basis'')
because we showed in Subsubsection~\ref{subsubsec_sifting} that
$\Sigma_1$ and $\Pi_4$ can be seen as uniformly and randomly chosen subsets
of $\Sigma$, conditioning on $F^\mathrm{sift'} = F^\mathrm{min} = \checkmark$.

We remark that this is not a straightforward application of Lemma~\ref{lemma_lln}
because the number $b = n_1 R'_+$ of rounds in $\overline{\Pi}$
is a \emph{random variable} and not a parameter.
Therefore, the computation is slightly more complicated
because all possible values of $R'_+ = r'_+$ need to be evaluated.
Nevertheless, using the condition
$R'_+ \ge \frac{1}{2T} - \nu$
and applying Lemma~\ref{lemma_lln} for any possible value of $R'_+$,
we are able to bound this probability and prove it exponentially~small.

Using the formulation of Lemma~\ref{lemma_lln}, we obtain
\vspace{-12pt}
\begin{adjustwidth}{-\extralength}{0cm}
\begin{eqnarray}
&&\pr \left[ R'_+ \ge \frac{1}{2T} - \nu ~~ \wedge ~~
\sum_{i = 1}^{k_4} 1\{V_i^4 \ne W_i^4\} \le k_4 \delta \right. \nonumber \\
&\wedge& \left. \sum_{i = 1}^{n_1} 1\{X'^1_i = 0 ~ \wedge ~ Y'^1_i = 1\} \ge
n_1 R'_+ \cdot (\delta + \nu) ~ \mid ~
F^\mathrm{sift'} = F^\mathrm{min} = \checkmark \right] \nonumber \\
&=& \pr \left[ R'_+ \ge \frac{1}{2T} - \nu
~~ \wedge ~~ \sum_{i = 1}^{k_4} 1\{V_i^4 = 0 ~ \wedge ~ W_i^4 = 1\} \le k_4 \delta
\right. \nonumber \\
&\wedge& \left. \sum_{i = 1}^{n_1} 1\{X'^1_i = 0 ~ \wedge ~ Y'^1_i = 1\} \ge
n_1 R'_+ \cdot (\delta + \nu) ~ \mid ~
F^\mathrm{sift'} = F^\mathrm{min} = \checkmark \right] \nonumber \\
&=& \sum_{j = \left\lceil n_1 \cdot \left( \frac{1}{2T} - \nu \right)
\right\rceil}^{n_1} \pr \left[ R'_+ = \frac{j}{n_1} ~~ \wedge ~~
\sum_{i = 1}^{k_4} 1\{V_i^4 = 0 ~ \wedge ~ W_i^4 = 1\} \le k_4 \delta
\right. \nonumber \\
&\wedge& \left. \sum_{i = 1}^{n_1} 1\{X'^1_i = 0 ~ \wedge ~ Y'^1_i = 1\} \ge
n_1 R'_+ \cdot (\delta + \nu) ~ \mid ~
F^\mathrm{sift'} = F^\mathrm{min} = \checkmark \right] \nonumber \\
&=& \sum_{j = \left\lceil n_1 \cdot \left( \frac{1}{2T} - \nu \right)
\right\rceil}^{n_1} \pr \left[ R'_+ = \frac{j}{n_1} ~ \mid ~
F^\mathrm{sift'} = F^\mathrm{min} = \checkmark \right] \cdot
\pr \left[ \sum_{i = 1}^{k_4} 1\{V_i^4 = 0 ~ \wedge ~ W_i^4 = 1\} \le k_4 \delta
\right. \nonumber \\
&\wedge& \left. \sum_{i = 1}^{n_1} 1\{X'^1_i = 0 ~ \wedge ~ Y'^1_i = 1\} \ge
n_1 R'_+ \cdot (\delta + \nu) ~ \mid ~ R'_+ = \frac{j}{n_1} ~ , ~
F^\mathrm{sift'} = F^\mathrm{min} = \checkmark \right].\label{eq_lln_p2}
\end{eqnarray}
\end{adjustwidth}
We can now bound this conditional probability, for each value of $j \in
\left[ n_1 \cdot \left( \frac{1}{2T} - \nu \right) ~ , ~ n_1 \right]$:
\begin{eqnarray}
&&\pr \left[ \sum_{i = 1}^{k_4} 1\{V_i^4 = 0 ~ \wedge ~ W_i^4 = 1\} \le k_4 \delta
\right. \nonumber \\
&\wedge& \left. \sum_{i = 1}^{n_1} 1\{X'^1_i = 0 ~ \wedge ~ Y'^1_i = 1\} \ge
n_1 R'_+ \cdot (\delta + \nu) ~ \mid ~ R'_+ = \frac{j}{n_1} ~ , ~
F^\mathrm{sift'} = F^\mathrm{min} = \checkmark \right] \nonumber \\
&=& \pr \left[ \sum_{i = 1}^{k_4} 1\{V_i^4 = 0 ~ \wedge ~ W_i^4 = 1\} \le
k_4 \delta \right. \nonumber \\
&\wedge& \left. \sum_{i = 1}^{n_1} 1\{X'^1_i = 0 ~ \wedge ~ Y'^1_i = 1\} \ge
j \cdot (\delta + \nu) ~ \mid ~ R'_+ = \frac{j}{n_1} ~ , ~
F^\mathrm{sift'} = F^\mathrm{min} = \checkmark \right] \nonumber \\
&=& \pr \left[ \sum_{i \in \Pi} Z_i \le k_4 \delta
~~ \wedge ~~ \sum_{i \in \overline{\Pi}} Z_i \ge j \cdot (\delta + \nu)
~ \mid ~ R'_+ = \frac{j}{n_1} ~ , ~
F^\mathrm{sift'} = F^\mathrm{min} = \checkmark \right] \nonumber \\
&\le& e^{-\frac{2 j k_4^2 \nu^2}{(k_4 + j)(k_4 + 1)}}
\le e^{-\frac{2 n_1 \cdot \left( \frac{1}{2T} - \nu \right) k_4^2
\nu^2}{\left(k_4 + n_1 \cdot \left( \frac{1}{2T} - \nu \right) \right)(k_4 +
1)}},\label{eq_lln_p2_cond}
\end{eqnarray}
where the last inequality results from the fact that
$j \ge n_1 \cdot \left( \frac{1}{2T} - \nu \right)$.

Substituting Equation~\eqref{eq_lln_p2_cond} into Equation~\eqref{eq_lln_p2},
we have
\begin{eqnarray}
&&\pr \left[ R'_+ \ge \frac{1}{2T} - \nu ~~ \wedge ~~
\sum_{i = 1}^{k_4} 1\{V_i^4 \ne W_i^4\} \le k_4 \delta \right. \nonumber \\
&\wedge& \left. \sum_{i = 1}^{n_1} 1\{X'^1_i = 0 ~ \wedge ~ Y'^1_i = 1\} \ge
n_1 R'_+ \cdot (\delta + \nu) ~ \mid ~
F^\mathrm{sift'} = F^\mathrm{min} = \checkmark \right] \nonumber \\
&=& \sum_{j = \left\lceil n_1 \cdot \left( \frac{1}{2T} - \nu \right)
\right\rceil}^{n_1} \pr \left[ R'_+ = \frac{j}{n_1} ~ \mid ~
F^\mathrm{sift'} = F^\mathrm{min} = \checkmark \right] \cdot
\pr \left[ \sum_{i = 1}^{k_4} 1\{V_i^4 = 0 ~ \wedge ~ W_i^4 = 1\} \le k_4 \delta
\right. \nonumber \\
&\wedge& \left. \sum_{i = 1}^{n_1} 1\{X'^1_i = 0 ~ \wedge ~ Y'^1_i = 1\} \ge
n_1 R'_+ \cdot (\delta + \nu) ~ \mid ~ R'_+ = \frac{j}{n_1} ~ , ~
F^\mathrm{sift'} = F^\mathrm{min} = \checkmark \right] \nonumber \\
&\le& \sum_{j = \left\lceil n_1 \cdot \left( \frac{1}{2T} - \nu \right)
\right\rceil}^{n_1} \pr \left[ R'_+ = \frac{j}{n_1} ~ \mid ~
F^\mathrm{sift'} = F^\mathrm{min} = \checkmark \right] \cdot
e^{-\frac{2 n_1 \cdot \left( \frac{1}{2T} - \nu \right) k_4^2
\nu^2}{\left(k_4 + n_1 \cdot \left( \frac{1}{2T} - \nu \right)\right)(k_4 + 1)}}
\nonumber \\
&\le& e^{-\frac{2 n_1 \cdot \left( \frac{1}{2T} - \nu \right) k_4^2
\nu^2}{\left(k_4 + n_1 \cdot \left( \frac{1}{2T} - \nu \right)\right)(k_4 + 1)}}.
\end{eqnarray}

\paragraph{\textbf{Bounding the third probability:}}
We need to bound
\vspace{-6pt}
\begin{adjustwidth}{-\extralength}{0cm}
\begin{equation}
\pr \left[ \sum_{i = 1}^{k_2} 1\{W_i^2 = 0\} \le k_2 \delta_\mathrm{mismatch}
~~ \wedge ~~ \sum_{i = 1}^{n_1} 1\{Y'^1_i = 0\} \ge
n_1 \cdot (\delta_\mathrm{mismatch} + \nu) ~ \mid ~
F^\mathrm{sift'} = F^\mathrm{min} = \checkmark \right],
\end{equation}
\end{adjustwidth}
where $W^2$ is Bob's substring corresponding to $\Pi_2$
(the $k_2$ TEST-Z-X rounds)
and $Y'^1$ is Bob's substring corresponding to $\Sigma_1$
(the $n_1$ INFO rounds) in the X-X hypothetical protocol.
Let us now define $X''^1$ and $Y''^1$ as Alice's and Bob's substrings
corresponding to $\Sigma_1$ (the $n_1$ INFO rounds)
\emph{in the Z-X hypothetical protocol}---namely,
\emph{assuming that Alice measured the INFO bits in the ``$z$ basis'',
and Bob measured the INFO bits in the ``$x$ basis''}.
We can notice that $Y'^1$ is completely identical to $Y''^1$,
because Bob's quantum operations (and Eve's attack) are completely independent
of Alice's basis choice for the INFO bits
(remembering that $D_i = 0$ for all INFO bits---namely, they are never discarded).
Therefore, $Y'^1 = Y''^1$, and the probability is
\begin{adjustwidth}{-\extralength}{0cm}
\begin{equation}
= \pr \left[ \sum_{i = 1}^{k_2} 1\{W_i^2 = 0\} \le k_2 \delta_\mathrm{mismatch}
~~ \wedge ~~ \sum_{i = 1}^{n_1} 1\{Y''^1_i = 0\} \ge
n_1 \cdot (\delta_\mathrm{mismatch} + \nu) ~ \mid ~
F^\mathrm{sift'} = F^\mathrm{min} = \checkmark \right].
\end{equation}
\end{adjustwidth}
We notice that all rates are evaluated in the Z-X hypothetical protocol;
that in all rounds, Bob measures in the ``$x$ basis''
(and Alice measures in the ``$z$ basis'');
and that Alice's measurement results are completely unconstrained
(namely, no discarding is possible, because $D_i = 0$ for all rounds
where Alice measures in the ``$z$ basis'').
We thus notice that the quantum behaviour of Alice, Bob, and Eve
is \emph{identical} on all these rounds in the Z-X hypothetical protocol.

Therefore, we can apply Lemma~\ref{lemma_lln} using the following parameters:
the random variables $Z = (Z_1, Z_2, \ldots, Z_N)$
represent the condition that Bob's bit is $0$
(namely, $Z_i$ is the value of $1 - W_i^2$ or $1 - Y''^1_i$, respectively);
the sampled subset $\Pi$ includes the $a = k_2$
TEST-Z-X rounds in the $\Pi_2$ subset chosen by the protocol,
and the rest $\overline{\Pi}$ includes the $b = n_1$ INFO rounds
in the $\Sigma_1$ subset chosen by the protocol
(note that Bob measures them in the ``$x$ basis'').
The sampled susbet $\Pi$ is completely independent of Bob's measurement results
(that are dictated solely by Eve's transmitted states and Alice's non-discarding
of the rounds)
because we showed in Subsubsection~\ref{subsubsec_sifting} that
$\Sigma_1$ and $\Pi_2$ can be seen as uniformly and randomly chosen subsets
of $\Sigma$, conditioning on $F^\mathrm{sift'} = F^\mathrm{min} = \checkmark$.
Using the formulation of Lemma~\ref{lemma_lln}, we obtain the following:
\vspace{-12pt}
\begin{adjustwidth}{-\extralength}{0cm}
\begin{eqnarray}
&&\pr \left[ \sum_{i = 1}^{k_2} 1\{W_i^2 = 0\} \le k_2 \delta_\mathrm{mismatch}
~~ \wedge ~~ \sum_{i = 1}^{n_1} 1\{Y'^1_i = 0\} \ge
n_1 \cdot (\delta_\mathrm{mismatch} + \nu) ~ \mid ~
F^\mathrm{sift'} = F^\mathrm{min} = \checkmark \right] \nonumber \\
&=& \pr \left[ \sum_{i = 1}^{k_2} 1\{W_i^2 = 0\} \le k_2 \delta_\mathrm{mismatch}
~~ \wedge ~~ \sum_{i = 1}^{n_1} 1\{Y''^1_i = 0\} \ge
n_1 \cdot (\delta_\mathrm{mismatch} + \nu) ~ \mid ~
F^\mathrm{sift'} = F^\mathrm{min} = \checkmark \right] \nonumber \\
&=& \pr \left[ \sum_{i \in \Pi} Z_i \le k_2 \delta_\mathrm{mismatch}
~~ \wedge ~~ \sum_{i \in \overline{\Pi}} Z_i \ge n_1 \cdot
(\delta_\mathrm{mismatch} + \nu) ~ \mid ~
F^\mathrm{sift'} = F^\mathrm{min} = \checkmark \right] \nonumber \\
&\le& e^{-\frac{2 n_1 k_2^2 \nu^2}{(k_2 + n_1)(k_2 + 1)}}.
\end{eqnarray}
\end{adjustwidth}

\paragraph{\textbf{Bounding the fourth probability:}}
We need to bound
\vspace{-12pt}
\begin{adjustwidth}{-\extralength}{0cm}
\begin{align}
\pr &\left[ R'_+ \ge \frac{1}{2T} - \nu ~~ \wedge ~~
\sum_{i = 1}^{n_1} 1\{X'^1_i = 0 ~ \wedge ~ Y'^1_i = 1\} \le
n_1 R'_+ \cdot (\delta + \nu) \right. \nonumber \\
&\wedge ~~ \left. \sum_{i = 1}^{n_1} 1\{Y'^1_i = 0\} \le
n_1 \cdot (\delta_\mathrm{mismatch} + \nu)
~~ \wedge ~~ \sum_{i = 1}^{n_1} 1\{X'^1_i \ne Y'^1_i\} > n_1 \delta'(\nu) ~ \mid ~
F^\mathrm{sift'} = F^\mathrm{min} = \checkmark \right],
\end{align}
\end{adjustwidth}
where $X'^1$ and $Y'^1$ are Alice's and Bob's substrings
corresponding to $\Sigma_1$ (the $n_1$ INFO rounds)
in the X-X hypothetical protocol,
and $R'_+$ is the ``$\xi_+$'' measurement rate of Alice among the INFO bits
in the X-X hypothetical protocol.
We prove this probability to be \emph{zero};
namely, we prove that these four conditions contradict each other
and cannot be all true.

Indeed, assume by contradiction that all four conditions hold:
\begin{eqnarray}
&&R'_+ \ge \frac{1}{2T} - \nu ~~ \wedge ~~
\sum_{i = 1}^{n_1} 1\{X'^1_i = 0 ~ \wedge ~ Y'^1_i = 1\} \le
n_1 R'_+ \cdot (\delta + \nu) \nonumber \\
&\wedge& \sum_{i = 1}^{n_1} 1\{Y'^1_i = 0\} \le
n_1 \cdot (\delta_\mathrm{mismatch} + \nu)
~~ \wedge ~~ \sum_{i = 1}^{n_1} 1\{X'^1_i \ne Y'^1_i\} > n_1 \delta'(\nu).
\end{eqnarray}
We can upper-bound $\sum_{i = 1}^{n_1} 1\{X'^1_i \ne Y'^1_i\}$
(which represents the total error rate
on the INFO bits in the X-X hypothetical protocol)
using the first three conditions, as well as the two following definitions:
$\delta'(\nu) \triangleq
\delta_\mathrm{mismatch} + \nu - \left(\frac{1}{2T} - \nu\right) \cdot
(1 - 2 \delta - 2 \nu)$
and $R'_+ \triangleq \frac{1}{n_1} \sum_{i = 1}^{n_1} 1\{X'^1_i = 0\}$. So:
\vspace{-12pt}
\begin{adjustwidth}{-\extralength}{0cm}
\begin{eqnarray}
\sum_{i = 1}^{n_1} 1\{X'^1_i \ne Y'^1_i\}
&=& \sum_{i = 1}^{n_1} 1\{X'^1_i = 0 ~ \wedge ~ Y'^1_i = 1\}
+ \sum_{i = 1}^{n_1} 1\{X'^1_i = 1 ~ \wedge ~ Y'^1_i = 0\} \nonumber \\
&=& \sum_{i = 1}^{n_1} 1\{X'^1_i = 0 ~ \wedge ~ Y'^1_i = 1\}
+ \sum_{i = 1}^{n_1} 1\{Y'^1_i = 0\}
- \sum_{i = 1}^{n_1} 1\{X'^1_i = 0 ~ \wedge ~ Y'^1_i = 0\} \nonumber \\
&=& \sum_{i = 1}^{n_1} 1\{X'^1_i = 0 ~ \wedge ~ Y'^1_i = 1\}
+ \sum_{i = 1}^{n_1} 1\{Y'^1_i = 0\} \nonumber \\
&-& \left[ \sum_{i = 1}^{n_1} 1\{X'^1_i = 0\}
- \sum_{i = 1}^{n_1} 1\{X'^1_i = 0 ~ \wedge ~ Y'^1_i = 1\} \right] \nonumber \\
&=& 2 \sum_{i = 1}^{n_1} 1\{X'^1_i = 0 ~ \wedge ~ Y'^1_i = 1\}
+ \sum_{i = 1}^{n_1} 1\{Y'^1_i = 0\}
- \sum_{i = 1}^{n_1} 1\{X'^1_i = 0\} \nonumber \\
&=& 2 \sum_{i = 1}^{n_1} 1\{X'^1_i = 0 ~ \wedge ~ Y'^1_i = 1\}
+ \sum_{i = 1}^{n_1} 1\{Y'^1_i = 0\}
- n_1 R'_+ \nonumber \\
&\le& 2 n_1 R'_+ \cdot (\delta + \nu)
+ n_1 \cdot (\delta_\mathrm{mismatch} + \nu)
- n_1 R'_+ \nonumber \\
&=& n_1 \cdot [2 R'_+ \cdot (\delta + \nu) + \delta_\mathrm{mismatch} + \nu - R'_+]
\nonumber \\
&=& n_1 \cdot [\delta_\mathrm{mismatch} + \nu - R'_+ \cdot (1 - 2 \delta - 2 \nu)]
\nonumber \\
&\le& n_1 \cdot \left[ \delta_\mathrm{mismatch} + \nu
- \left( \frac{1}{2T} - \nu \right) \cdot (1 - 2 \delta - 2 \nu) \right]
= n_1 \delta'(\nu),
\end{eqnarray}
\end{adjustwidth}
which strictly contradicts the fourth condition.
(In the last inequality, we also used the condition $\nu \le \frac{1}{2} - \delta$,
which means that $1 - 2 \delta - 2 \nu \ge 0$.)
Therefore, our probability is $0$.

\paragraph{\textbf{Summary of the proof:}}
Combining our four bounds, we obtain the following inequality:
\vspace{-12pt}
\begin{adjustwidth}{-\extralength}{0cm}
\begin{eqnarray}
&&\pr \left[ F^\mathrm{pe} = \checkmark ~ \wedge ~ \Omega_0 ~ \mid ~
F^\mathrm{sift'} = F^\mathrm{min} = \checkmark \right] \nonumber \\
&\le& \pr \left[ R'_+ \le \frac{1}{2T} - \nu ~ \mid ~
F^\mathrm{sift'} = F^\mathrm{min} = \checkmark \right] \nonumber \\
&+& \pr \left[ R'_+ \ge \frac{1}{2T} - \nu ~~ \wedge ~~
\sum_{i = 1}^{k_4} 1\{V_i^4 \ne W_i^4\} \le k_4 \delta \right. \nonumber \\
&\wedge& \left. \sum_{i = 1}^{n_1} 1\{X'^1_i = 0 ~ \wedge ~ Y'^1_i = 1\} \ge
n_1 R'_+ \cdot (\delta + \nu) ~ \mid ~
F^\mathrm{sift'} = F^\mathrm{min} = \checkmark \right] \nonumber \\
&+& \pr \left[ \sum_{i = 1}^{k_2} 1\{W_i^2 = 0\} \le k_2 \delta_\mathrm{mismatch}
~~ \wedge ~~ \sum_{i = 1}^{n_1} 1\{Y'^1_i = 0\} \ge
n_1 \cdot (\delta_\mathrm{mismatch} + \nu) ~ \mid ~
F^\mathrm{sift'} = F^\mathrm{min} = \checkmark \right] \nonumber \\
&+& \pr \left[ R'_+ \ge \frac{1}{2T} - \nu ~~ \wedge ~~
\sum_{i = 1}^{n_1} 1\{X'^1_i = 0 ~ \wedge ~ Y'^1_i = 1\} \le
n_1 R'_+ \cdot (\delta + \nu) \right. \nonumber \\
&\wedge& \left. \sum_{i = 1}^{n_1} 1\{Y'^1_i = 0\} \le
n_1 \cdot (\delta_\mathrm{mismatch} + \nu)
~~ \wedge ~~ \sum_{i = 1}^{n_1} 1\{X'^1_i \ne Y'^1_i\} > n_1 \delta'(\nu) ~ \mid ~
F^\mathrm{sift'} = F^\mathrm{min} = \checkmark \right] \nonumber \\
&\le& e^{-2 n_1 \nu^2}
+ e^{-\frac{2 n_1 \cdot \left( \frac{1}{2T} - \nu \right) k_4^2
\nu^2}{\left(k_4 + n_1 \cdot \left( \frac{1}{2T} - \nu \right)\right)(k_4 + 1)}}
+ e^{-\frac{2 n_1 k_2^2 \nu^2}{(k_2 + n_1)(k_2 + 1)}} + 0 \nonumber \\
&=& \epsilon(\nu)^2.
\end{eqnarray}
\end{adjustwidth}
The rest of the proof is identical to the proof of Proposition~8
in Subsection~6.3 of~\cite{TL17}, using our parameters
$\epsilon(\nu), X'^1, Y'^1, n_1, \Omega_0, \delta'(\nu)$
and conditioning all probabilities and entropies
on $F^\mathrm{sift'} = F^\mathrm{min} = \checkmark$.
(A small algebraic difference is that our set $\Omega_0$ requires
the strong inequality $1 \left\{ \sum_{i = 1}^{n_1} 1\{ X'^1_i \ne Y'^1_i \} >
n_1 \delta'(\nu) \right\}$, while $\Omega$'s definition in~\cite{TL17}
only requires a weak inequality ($1 \left\{ \sum_{i = 1}^n 1\{ X_i \ne Y_i \} \ge
n \cdot (\delta + \nu) \right\}$), but the proof still holds.)
Therefore, we obtain the following:
\begin{equation}
H_\mathrm{max}^{\epsilon(\nu)}(\sX'^1 \wedge \sF^\mathrm{pe} = \checkmark \mid
\sY'^1 ~ , ~ \sF^\mathrm{sift'} = \sF^\mathrm{min} = \checkmark)_{\sigma'}
\le n_1 \cdot h_2(\delta'(\nu)),
\end{equation}
as we wanted.
\end{proof}

\subsubsection{\label{subsubsec_eb_conc}Security Theorem
for the Modified Entanglement-Based Protocol}
\textls[-25]{Applying the entire proof described in Section~6 of~\cite{TL17} to
our modified entanglement-based protocol described in Subsection~\ref{subsec_eb},
with the modifications described in
Subsubsections~\ref{subsubsec_sifting}--\ref{subsubsec_lln},}
yields the following security result:
\begin{Corollary}\label{corollary_security_eb}
For the modified entanglement-based protocol
described in Subsection~\ref{subsec_eb}, we denote the final state
as $\omega_{\sK_\sA \sK_\sB \sS \sC \sF \sE}$,
where $\sK_\sA$ is the final key generated by Alice
and $\sK_\sB$ is the final key generated by Bob (both consisting of $\ell$ bits),
$\sE$ is Eve's ancillary quantum system,
and $\sS, \sC, \sF$ consist of information published by Alice and Bob
(where $H_\mathrm{ec}, H_\mathrm{pa}, Z, T, F^\mathrm{ec}, r, t$ are used
in the error correction and privacy amplification steps elaborated in~\cite{TL17}):
\begin{eqnarray}
\sS &\triangleq& (\Phi_\sA, \Phi_\sB, \Pi_1, \Pi_2, \Pi_3, \Pi_4, \Sigma_1,
H_\mathrm{ec}, H_\mathrm{pa}), \\
\sC &\triangleq& (D, \Sigma, V^1, V^2, V^3, V^4, Z, T), \\
\sF &\triangleq& (F^\mathrm{sift'}, F^\mathrm{min}, F^\mathrm{pe}, F^\mathrm{ec}).
\end{eqnarray}
We also denote $\omega_\sU \triangleq \frac{1}{2^\ell} \sum_{k \in \{0, 1\}^\ell}
\ket{k}_{\sK_\sA} \bra{k}_{\sK_\sA} \otimes \ket{k}_{\sK_\sB} \bra{k}_{\sK_\sB}$
(an ideal key: a uniformly random final key, identical for Alice and Bob)
and $\omega_{\sS \sC \sF \sE} \triangleq \tr_{\sK_\sA \sK_\sB}
(\omega_{\sK_\sA \sK_\sB \sS \sC \sF \sE})$.
It then holds that
\vspace{-12pt}
\begin{adjustwidth}{-\extralength}{0cm}
\begin{eqnarray}
&&\frac{1}{2} \tr \left| \omega_{\sK_\sA \sK_\sB \sS \sC \sF \sE ~ \wedge ~
\sF^\mathrm{pe} = \sF^\mathrm{ec} = \checkmark ~ \mid ~
\sF^\mathrm{sift'} = \sF^\mathrm{min} = \checkmark}
- \omega_\sU \otimes \omega_{\sS \sC \sF \sE ~ \wedge ~
\sF^\mathrm{pe} = \sF^\mathrm{ec} = \checkmark ~ \mid ~
\sF^\mathrm{sift'} = \sF^\mathrm{min} = \checkmark} \right| \nonumber \\
&\le& 2^{-t} + \inf_{\nu ~ \mid ~ 0 < \nu < \frac{1}{2} - \delta ~ , ~
0 < \delta'(\nu) < \frac{1}{2}}
(\epsilon_\mathrm{pa}(\nu) + \epsilon_\mathrm{pe}(\nu)),
\end{eqnarray}
\end{adjustwidth}
for any possible attack by Eve, where we define
\begin{eqnarray}
\epsilon_\mathrm{pa}(\nu) &\triangleq& \frac{1}{2} \sqrt{2^{-n_1 \cdot
\left[ \log_2 \left( \frac{1}{c} \right) - h_2(\delta'(\nu)) \right] +
r + t + \ell}}, \\
\epsilon_\mathrm{pe}(\nu) &\triangleq& 2 \epsilon(\nu), \\
\delta'(\nu) &\triangleq& \delta_\mathrm{mismatch} + \nu
- \left(\frac{1}{2T} - \nu\right) \cdot (1 - 2 \delta - 2 \nu), \\
c &\triangleq& \max\left(|\braket{0}{\xi_+}_{\sA_i}|^2,
|\braket{0}{\xi_-}_{\sA_i}|^2,
|\braket{1}{\xi_+}_{\sA_i}|^2, |\braket{1}{\xi_-}_{\sA_i}|^2\right), \\
\epsilon(\nu) &\triangleq& \sqrt{e^{-2 n_1 \nu^2}
+ e^{-\frac{2 n_1 \cdot \left( \frac{1}{2T} - \nu \right) k_4^2
\nu^2}{\left(k_4 + n_1 \cdot \left( \frac{1}{2T} - \nu \right)\right)(k_4 + 1)}}
+ e^{-\frac{2 n_1 k_2^2 \nu^2}{(k_2 + n_1)(k_2 + 1)}}}, \\
h_2(x) &\triangleq& -x \log_2(x) - (1-x) \log_2(1-x).
\end{eqnarray}
\end{Corollary}
This result is similar to Theorems~2--3 in Section~5 of~\cite{TL17}
but has different parameters.
According to standard definitions of composable security for QKD
(e.g.,~\cite{renner_thesis08}),
this proves security of the modified entanglement-based protocol
and gives a tight finite-key rate.

\subsection{\label{subsec_eb_red}Reduction of the Original Protocol
to the Modified Entanglement-Based Protocol}
Intuitively, to prove security of our original prepare-and-measure protocol
from Section~\ref{sec_def}, we perform a reduction to the entanglement-based
protocol: namely, we show that the modified entanglement-based protocol includes
the prepare-and-measure protocol as a special case.
More precisely, for any possible attack of Eve on the prepare-and-measure protocol,
we need to show that there exists an equivalent attack on the modified
entanglement-based protocol leading to the same output for both protocols.

The proof intuitively works as follows: in the prepare-and-measure protocol,
instead of generating one of the states
$\{\ket{\gamma_0}_{\sB_i}, \ket{\gamma_1}_{\sB_i}, \ket{\gamma_+}_{\sB_i}\}$,
Alice can instead generate the following entangled state:
\begin{equation}
\ket{\Psi}_{\sA_i\sB_i} \triangleq \frac{\ket{0}_{\sA_i} \ket{\gamma_0}_{\sB_i}
+ \ket{1}_{\sA_i} \ket{\gamma_1}_{\sB_i}}{\sqrt{2}}
= \frac{\ket{\xi_+}_{\sA_i} \ket{\gamma_+}_{\sB_i}
+ \sqrt{2T - 1} \ket{\xi_-}_{\sA_i}
\ket{\gamma_-}_{\sB_i}}{\sqrt{2T}},\label{eq:psi_ent}
\end{equation}
where the equality between the two expressions in Equation~\eqref{eq:psi_ent}
can be shown algebraically using Equations~\eqref{eq:gamma_+}--\eqref{eq:gamma_-}
and~\eqref{eq:xi_+}--\eqref{eq:xi_-}.
Then, Alice either measures her subsystem $\sA_i$
in the standard (``$z$'') basis $\{\ket{0}_{\sA_i}, \ket{1}_{\sA_i}\}$
with probability $p'^\sA_z$, or measures it
in the conjugate (``$x$'') basis $\{\ket{\xi_+}_{\sA_i}, \ket{\xi_-}_{\sA_i}\}$
with probability $p'^\sA_x$. Either way, she sends the resulting state
in subsystem $\sB_i$ to Bob (immediately notifying Bob and cancelling the round
if she measured ``$\xi_-$'' in the ``$x$ basis'').
This procedure is equivalent to our original prepare-and-measure protocol,
but it works within the framework of the modified entanglement-based protocol
(assuming Alice measures and discards the round \emph{before} she sends Bob
his part of the state), which proves the reduction.

Formally, we use an adapted version of the reduction in Section~9 of~\cite{TL17}.
First, given the parameters of the original prepare-and-measure protocol
(described in its Step~\ref{pm_step_1}), we must define all the parameters of
the modified entanglement-based protocol (described in its Step~\ref{eb_step_1}), as follows:
\begin{itemize}
\item The parameters $\ket{\gamma_0}_{\sB_i}, \ket{\gamma_1}_{\sB_i},
\ket{\gamma_+}_{\sB_i}, \{M^{\sZ,t}_{\sB_i}\}_{t \in \{0,1\}},
\{M^{\sX,t}_{\sB_i}\}_{t \in \{0,1\}}, m, k_1, k_2, k_3, k_4, n_1, \delta$,
and $\delta_\mathrm{mismatch}$ are all identical for the two protocols.
(From $\ket{\gamma_0}_{\sB_i}, \ket{\gamma_1}_{\sB_i}, \ket{\gamma_+}_{\sB_i}$
we can infer $a$, $b$, and $T \triangleq |a|^2 + |b|^2$.)
The error correction and privacy amplification parameters
(from~\cite{TL17}) are also identical in both protocols.
\item Given the parameters $m, T$,
we choose the parameter $M'$ of the modified entanglement-based protocol to be
\begin{equation}
M' = \frac{m}{\frac{1}{2T} - \nu_0},
\end{equation}
where $0 < \nu_0 < \frac{1}{2T}$ is chosen \emph{freely},
without any constraint,
to reach the desired trade-off between performance (number of needed rounds)
and robustness (success probability of the sifting procedure).
\item Given the parameters $p^\sA_z, p^\sA_x$
of the prepare-and-measure protocol and the parameter $T$,
we choose the parameters $p'^\sA_z, p'^\sA_x$
of the modified entanglement-based protocol to~be
\begin{equation}\label{eq:pm_to_eb_pA}
p'^\sA_z = \frac{p^\sA_z}{p^\sA_z + 2T p^\sA_x} ~~~ , ~~~
p'^\sA_x = \frac{2T p^\sA_x}{p^\sA_z + 2T p^\sA_x} ~~~ .
\end{equation}
\item Given the parameters $p^\sB_z, p^\sB_x$
of the prepare-and-measure protocol,
we choose the parameters $p'^\sB_z, p'^\sB_x$
of the modified entanglement-based protocol to be
\begin{equation}\label{eq:pm_to_eb_pB}
p'^\sB_z = p^\sB_z ~~~ , ~~~ p'^\sB_x = p^\sB_x ~~~ .
\end{equation}
\end{itemize}

Using these parameters, it is easy to verify that the output of
the prepare-and-measure protocol
(conditioned on $F^\mathrm{min} = \checkmark$)
is identical to the output of the modified entanglement-based protocol
(conditioned on $F^\mathrm{sift'} = F^\mathrm{min} = \checkmark$)
if Eve performs the same attack
on the first $m$ non-discarded rounds in both protocols.
Formally, the differences between the protocols are settled as follows:
\begin{itemize}
\item The modified entanglement-based protocol includes the possibility of
\emph{discarded rounds} (where Alice measures ``$\xi_-$'')
which are immediately notified to Bob and Eve,
while the prepare-and-measure protocol does not allow them.
For this, we use the explanation in Subsubsection~\ref{subsubsec_sifting}
to divide the process into two stages (again, this division works with respect
to the probability distribution, \emph{not} to the actual quantum operations):
stage~\ref{sifting_stage_1}, where Alice determines which rounds are
discarded; and stage~\ref{sifting_stage_2}, where Alice determines the basis
for measuring all the non-discarded rounds.
As explained in Subsubsection~\ref{subsubsec_sifting},
stage~\ref{sifting_stage_1} is in fact independent of the bases
used for the non-discarded rounds; furthermore, the results
of stage~\ref{sifting_stage_1} are promptly communicated to Eve,
who can devise her attack accordingly.
Meanwhile, stage~\ref{sifting_stage_2} is completely identical between
the two protocols, as shown in the next item.
\item \emph{Alice's preparation} is different between the two protocols:

In the prepare-and-measure protocol,
Alice randomly chooses $\Phi_\sA \in \{0, 1\}^m$
(where each bit, independently, is $0$ with probability $p^\sA_z$
or $1$ with probability $p^\sA_x$)
and chooses $R \in \{0, 1\}^m$ uniformly at random,
which lead to the preparation of $\ket{\gamma_0}_{\sB_i}$,
$\ket{\gamma_1}_{\sB_i}$, or $\ket{\gamma_+}_{\sB_i}$ with probabilities
$\frac{p^\sA_z}{2}$, $\frac{p^\sA_z}{2}$, and $p^\sA_x$, respectively.
These probabilities are independent between the rounds.

In the modified entanglement-based protocol, Alice generates the following
state $\ket{\Psi}_{\sA_i \sB_i}$ for each round $i$:
\begin{equation}
\ket{\Psi}_{\sA_i \sB_i} \triangleq \frac{\ket{0}_{\sA_i} \ket{\gamma_0}_{\sB_i}
+ \ket{1}_{\sA_i} \ket{\gamma_1}_{\sB_i}}{\sqrt{2}}
= \frac{\ket{\xi_+}_{\sA_i} \ket{\gamma_+}_{\sB_i}
+ \sqrt{2T - 1} \ket{\xi_-}_{\sA_i} \ket{\gamma_-}_{\sB_i}}{\sqrt{2T}},
\end{equation}
randomly chooses the measurement bases $\Phi_\sA \in \{0, 1\}^{M'}$
(where each bit, independently, is $0$ with probability $p'^\sA_z$
or $1$ with probability $p'^\sA_x$), performs the measurement,
publicly discards the round if she obtains ``$\xi_-$'',
and keeps the result secret otherwise.
(In fact, Alice's measurement is delayed to Step~\ref{eb_step_a_mes}
if the chosen basis is the ``$z$ basis'',
as described in Subsection~\ref{subsec_eb}.)

As explained in Subsubsection~\ref{subsubsec_sifting}
(Equations~\eqref{eq:eb_phiA_0} and~\eqref{eq:eb_phiA_1}), for each
\emph{non-discarded} round in the modified entanglement-based protocol,
the probabilities that Alice measures
``$0$'', ``$1$'', or ``$\xi_+$'' (leading to her sending to Bob
$\ket{\gamma_0}_{\sB_i}$, $\ket{\gamma_1}_{\sB_i}$, or $\ket{\gamma_+}_{\sB_i}$,
respectively) are
\begin{equation}
\pr_{\sA_i}(0) = \pr_{\sA_i}(1) =
\frac{1}{2} \cdot \frac{p'^\sA_z}{p'^\sA_z + \frac{p'^\sA_x}{2T}} ~~~ , ~~~
\pr_{\sA_i}(\xi_+) = \frac{1}{2T} \cdot
\frac{p'^\sA_x}{p'^\sA_z + \frac{p'^\sA_x}{2T}} ~~~ .
\end{equation}
\textls[-15]{Substituting Equation~\eqref{eq:pm_to_eb_pA}
(and the fact $p^\sA_z + p^\sA_x = 1$), we obtain the following probabilities:}
\begin{eqnarray}
\pr_{\sA_i}(0) = \pr_{\sA_i}(1) &=& \frac{1}{2} \cdot
\frac{\frac{p^\sA_z}{p^\sA_z + 2T p^\sA_x}}{\frac{p^\sA_z}{p^\sA_z + 2T p^\sA_x}
+ \frac{\frac{2T p^\sA_x}{p^\sA_z + 2T p^\sA_x}}{2T}}
= \frac{1}{2} \cdot \frac{p^\sA_z}{p^\sA_z + p^\sA_x} = \frac{p^\sA_z}{2}, \\
\pr_{\sA_i}(\xi_+) &=& \frac{1}{2T} \cdot
\frac{\frac{2T p^\sA_x}{p^\sA_z + 2T p^\sA_x}}{\frac{p^\sA_z}{p^\sA_z + 2T p^\sA_x}
+ \frac{\frac{2T p^\sA_x}{p^\sA_z + 2T p^\sA_x}}{2T}}
= \frac{1}{2T} \cdot \frac{2T p^\sA_x}{p^\sA_z + p^\sA_x} = p^\sA_x,
\end{eqnarray}
which are independent between the rounds and
identical to the prepare-and-measure probabilities found above.
Therefore, Alice's preparation results are \emph{identical}
on the (non-discarded) rounds of both protocols,
even when conditioning on $F^\mathrm{sift'} = \checkmark$
in the modified entanglement-based protocol.
\item \emph{Eve's attack} is slightly different between the two protocols:
on the prepare-and-measure protocol, it is applied to the $m$ rounds
which are all relevant,
while on the modified entanglement-based protocol, it is applied to all $M'$ rounds
(including the discarded rounds) when Eve knows \emph{ahead of time} which
rounds are discarded.

We need to prove that any attack that Eve applies
to the $m$ rounds of the prepare-and-measure protocol
can also be applied to the relevant rounds of the modified entanglement-based
(namely, to the $m$ rounds in $\Sigma$,
which are the first $m$ rounds not discarded by Alice).
This is indeed true because in the modified entanglement-based protocol,
Eve knows ahead of time (before she applies her attack) which rounds are discarded,
and therefore, she knows exactly which rounds are included in $\Sigma$
and can apply her attack only to them.
This means that any attack by Eve on the $m$ rounds of the prepare-and-measure
protocol is a completely legitimate and valid attack on the $m$ rounds in $\Sigma$
of the modified entanglement-based protocol, and it gives the same outputs
in both protocols.
\item The rest of the steps in the prepare-and-measure protocol
(Steps~\ref{pm_step_min}--\ref{pm_step_ec_pa})
are identical to the rest of the steps
in the modified entanglement-based protocol
(Steps~\ref{eb_step_min}--\ref{eb_step_ec_pa}),
except the delayed measurement in Steps~\ref{eb_step_a_mes}
and~\ref{eb_step_b_mes} of the modified entanglement-based~protocol.
\end{itemize}

From the above, we can deduce that any attack by Eve on the prepare-and-measure
protocol can also be applied to the modified entanglement-based protocol,
giving exactly the same output.
This conclusion only applies when we condition on
$F^\mathrm{min} = \checkmark$
(for the prepare-and-measure protocol)
and $F^\mathrm{sift'} = F^\mathrm{min} = \checkmark$
(for the modified entanglement-based protocol), which is indeed the case
in our security proof in Subsection~\ref{subsec_security}.

We therefore obtain the following result:
\begin{Corollary}\label{corollary_security_red}
If the modified entanglement-based protocol is secure with
a specific security parameter $\epsilon$,
the prepare-and-measure protocol is secure with the same security parameter.
\end{Corollary}

Combining Corollaries~\ref{corollary_security_eb} and~\ref{corollary_security_red},
we obtain the final security result for the prepare-and-measure protocol:
\begin{Corollary}\label{corollary_security_pm}
For the prepare-and-measure protocol
described in Section~\ref{sec_def}, we denote the final state
as $\omega_{\sK_\sA \sK_\sB \sS \sC \sF \sE}$,
where $\sK_\sA$ is the final key generated by Alice
and $\sK_\sB$ is the final key generated by Bob (both consisting of $\ell$ bits),
$\sE$ is Eve's ancillary quantum system,
and $\sS, \sC, \sF$ consist of information published by Alice and Bob
(where $H_\mathrm{ec}, H_\mathrm{pa}, Z, T, F^\mathrm{ec}, r, t$ are used
in the error correction and privacy amplification steps elaborated in~\cite{TL17}):
\begin{eqnarray}
\sS &\triangleq& (\Phi_\sA, \Phi_\sB, \Pi_1, \Pi_2, \Pi_3, \Pi_4, \Sigma_1,
H_\mathrm{ec}, H_\mathrm{pa}), \\
\sC &\triangleq& (V^1, V^2, V^3, V^4, Z, T), \\
\sF &\triangleq& (F^\mathrm{min}, F^\mathrm{pe}, F^\mathrm{ec}).
\end{eqnarray}
We also denote $\omega_\sU \triangleq \frac{1}{2^\ell} \sum_{k \in \{0, 1\}^\ell}
\ket{k}_{\sK_\sA} \bra{k}_{\sK_\sA} \otimes \ket{k}_{\sK_\sB} \bra{k}_{\sK_\sB}$
(an ideal key: a uniformly random final key, identical for Alice and Bob)
and $\omega_{\sS \sC \sF \sE} \triangleq \tr_{\sK_\sA \sK_\sB}
(\omega_{\sK_\sA \sK_\sB \sS \sC \sF \sE})$.
It then holds that
\begin{eqnarray}
&&\frac{1}{2} \tr \left| \omega_{\sK_\sA \sK_\sB \sS \sC \sF \sE ~ \wedge ~
\sF^\mathrm{pe} = \sF^\mathrm{ec} = \checkmark ~ \mid ~
\sF^\mathrm{min} = \checkmark}
- \omega_\sU \otimes \omega_{\sS \sC \sF \sE ~ \wedge ~
\sF^\mathrm{pe} = \sF^\mathrm{ec} = \checkmark ~ \mid ~
\sF^\mathrm{min} = \checkmark} \right| \nonumber \\
&\le& 2^{-t} + \inf_{\nu ~ \mid ~ 0 < \nu < \frac{1}{2} - \delta ~ , ~
0 < \delta'(\nu) < \frac{1}{2}}
(\epsilon_\mathrm{pa}(\nu) + \epsilon_\mathrm{pe}(\nu)),
\end{eqnarray}
for any possible attack by Eve, where we define
\begin{eqnarray}
\epsilon_\mathrm{pa}(\nu) &\triangleq& \frac{1}{2} \sqrt{2^{-n_1 \cdot
\left[ \log_2 \left( \frac{1}{c} \right) - h_2(\delta'(\nu)) \right] +
r + t + \ell}}, \\
\epsilon_\mathrm{pe}(\nu) &\triangleq& 2 \epsilon(\nu), \\
\delta'(\nu) &\triangleq& \delta_\mathrm{mismatch} + \nu
- \left(\frac{1}{2T} - \nu\right) \cdot (1 - 2 \delta - 2 \nu), \\
c &\triangleq& \max\left(|\braket{0}{\xi_+}_{\sA_i}|^2,
|\braket{0}{\xi_-}_{\sA_i}|^2,
|\braket{1}{\xi_+}_{\sA_i}|^2, |\braket{1}{\xi_-}_{\sA_i}|^2\right), \\
\epsilon(\nu) &\triangleq& \sqrt{e^{-2 n_1 \nu^2}
+ e^{-\frac{2 n_1 \cdot \left( \frac{1}{2T} - \nu \right) k_4^2
\nu^2}{\left(k_4 + n_1 \cdot \left( \frac{1}{2T} - \nu \right)\right)(k_4 + 1)}}
+ e^{-\frac{2 n_1 k_2^2 \nu^2}{(k_2 + n_1)(k_2 + 1)}}}, \\
h_2(x) &\triangleq& -x \log_2(x) - (1-x) \log_2(1-x).
\end{eqnarray}
\end{Corollary}

\section{\label{sec_rest}Necessity of the Restriction to Three Source States}
In our protocol, similarly to the ``loss tolerant''
protocol~\cite{lt_qkd14,lt_qkd_exp15,lt_qkd_finite_key15,lt_qkd_gen19},
only three source states are used.
This restriction is necessary in the imperfect-generation regime,
as we briefly explain below.

Let us assume that our protocol emits \emph{four} input states (similarly to BB84),
denoted $\ket{\gamma_0}, \ket{\gamma_1}, \ket{\gamma_+}, \ket{\gamma_-}$.
For standard security analysis to work, the following conditions is required
for some $0 < p < 1$ and $0 < q < 1$:
\begin{equation}\label{eq:necess_cond}
p \ket{\gamma_0} \bra{\gamma_0} + (1-p) \ket{\gamma_1} \bra{\gamma_1}
= q \ket{\gamma_+} \bra{\gamma_+} + (1-q) \ket{\gamma_-} \bra{\gamma_-},
\end{equation}
which means that Alice sends to Bob \emph{identical} mixed states
in each round of the protocol, independently of the chosen basis.
(Otherwise, Eve may gain information on the basis and attack differently
on each basis, which refutes the crucial possibility of comparing her attack's
influence on different bases.)

For meeting the above condition,
we obviously need $\ket{\gamma_+}$ and $\ket{\gamma_-}$ to be in the
two-dimensional Hilbert subspace spanned by $\ket{\gamma_0}$ and $\ket{\gamma_1}$.
Therefore, we require (for some $a, b, c, d \in \mathbb{C}$):
\begin{equation}\label{eq:necess_gamma}
\ket{\gamma_+} = a \ket{\gamma_0} + b \ket{\gamma_1} ~~~ , ~~~
\ket{\gamma_-} = c \ket{\gamma_0} + d \ket{\gamma_1}.
\end{equation}
Substituting this into Equation~\eqref{eq:necess_cond}, we obtain the following:
\vspace{-12pt}
\begin{adjustwidth}{-\extralength}{0cm}
\begin{eqnarray}
q \ket{\gamma_+} \bra{\gamma_+} + (1-q) \ket{\gamma_-} \bra{\gamma_-}
&=& q \cdot [a \ket{\gamma_0} + b \ket{\gamma_1}]
\cdot [a^\star \bra{\gamma_0} + b^\star \bra{\gamma_1}] \nonumber \\
&+& (1-q) \cdot [c \ket{\gamma_0} + d \ket{\gamma_1}]
\cdot [c^\star \bra{\gamma_0} + d^\star \bra{\gamma_1}] \nonumber \\
&=& [q \cdot |a|^2 + (1-q) \cdot |c|^2] \cdot \ket{\gamma_0}\bra{\gamma_0}
+ [q \cdot a b^\star + (1-q) \cdot c d^\star] \cdot \ket{\gamma_0}\bra{\gamma_1}
\nonumber \\
&+& [q \cdot a^\star b + (1-q) \cdot c^\star d] \cdot \ket{\gamma_1}\bra{\gamma_0}
+ [q \cdot |b|^2 + (1-q) \cdot |d|^2] \cdot \ket{\gamma_1}\bra{\gamma_1}. ~~~~~~~
\label{eq:necess_cond_x}
\end{eqnarray}
\end{adjustwidth}
We thus obtain the following conditions for equality between
Equations~\eqref{eq:necess_cond} and \eqref{eq:necess_cond_x}:
\begin{eqnarray}
q \cdot |a|^2 + (1-q) \cdot |c|^2 &=& p,\label{eq:necess_cond_1}\\
q \cdot |b|^2 + (1-q) \cdot |d|^2 &=& 1-p,\label{eq:necess_cond_2}\\
q \cdot a b^\star + (1-q) \cdot c d^\star &=& 0,\label{eq:necess_cond_3}\\
q \cdot a^\star b + (1-q) \cdot c^\star d &=& 0.\label{eq:necess_cond_4}
\end{eqnarray}
\textls[-15]{The two last equations are the complex conjugates of one another,
so one of them is sufficient.}

Therefore, for standard security proofs to work, we require very stringent
conditions on $a,b,c,d$. In particular,
according to Equation~\eqref{eq:necess_cond_4}, we require
\begin{equation}
q = -\frac{c^\star d}{a^\star b - c^\star d}
= \frac{c^\star d}{c^\star d - a^\star b},
\end{equation}
and for $q$ to be real (and satisfy $0 < q < 1$), the complex phases of
$c^\star d$ and $a^\star b$ must be \emph{opposite}
(namely, they must differ by $\pm \pi$,
which is equivalent to having opposite signs).

This requirement seriously restricts the possible values on
$\ket{\gamma_0}, \ket{\gamma_1}, \ket{\gamma_+}, \ket{\gamma_-}$.
In particular, if we assume (without loss of generality) that $a$ and $c$
are real and non-negative,
it requires $b$ and $d$ to have opposite phases. Namely,
\begin{equation}
\ket{\gamma_+} = |a| \ket{\gamma_0} + |b| e^{i \phi} \ket{\gamma_1} ~~~ , ~~~
\ket{\gamma_-} = |c| \ket{\gamma_0} - |d| e^{i \phi} \ket{\gamma_1},
\end{equation}
where $|c|$ and $|d|$ are dictated by $|a|$ and $|b|$, respectively
(see Equations~\eqref{eq:necess_cond_1} and \eqref{eq:necess_cond_2}).

The above analysis means that $\ket{\gamma_-}$ is, in fact, completely determined
by the choice of $\ket{\gamma_0}, \ket{\gamma_1}, \ket{\gamma_+}$
(because $|c|$, $|d|$, and $\phi$ can all be inferred from $\ket{\gamma_+}$).
From a realistic point of view, this means that a four-state protocol
measured with two bases could be \emph{practically insecure} whenever
a slight deviation of $\ket{\gamma_-}$ (or of the states
$\ket{\gamma_0}, \ket{\gamma_1}, \ket{\gamma_+}$ which determine it)
causes the protocol to violate the conditions
of Equations~\eqref{eq:necess_cond_1}--\eqref{eq:necess_cond_4}.
Essentially, this means that in the presence of source imperfections,
the use of at most three states (or, alternatively,
measurements in three or more bases, which we do not explore here)
is \emph{required} for practical security,
and the use of four states could lead to practical security issues.

\section{\label{sec_conc}Conclusions}
To sum up, we have found a new way to analyse
the security of practical QKD protocols
by generalizing the results of~\cite{TL17} to more practical protocols
(using a modified entropic uncertainty relation and
a refined analysis of finite-key statistics).
Our proof, compared with other proofs, is rigorous, careful,
and simple, aiming to make it easy-to-use in the lossless qubit regime
(its extension to losses and decoy states is left for future research
because they present specific hurdles in this analysis regime:
in particular, losses would need to be declared by Eve
in the modified entanglement-based protocol, which could complicate the analysis).
We believe that our suggested tools can contribute to benchmarking and certifying
the security of practical implementations of~QKD.

\vspace{6pt}

\authorcontributions{Conceptualization, G.B.  and R.L.;
physical modelling, G.B.,  N.G., R.L., and S.V.;
formal analysis, M.B.  and R.L.;
writing---original draft preparation, R.L.;
writing---review and editing, G.B.,  R.L., and S.V.
All authors have read and agreed to the published version of the~manuscript.}

\funding{The work of G.B.\ was supported in part by
Canada's Natural Sciences and Engineering Research Council (NSERC),
Qu\'ebec's Institut transdisciplinaire d'information quantique \mbox{(INTRIQ)},
and the Canada Research Chair Program.
The work of N.G.\ and S.V.\ was supported in part by NSERC and INTRIQ.
The work of R.L.\ was supported in part by the Canada Research Chair Program,
the Technion's Helen Diller Quantum Center (Haifa, Israel),
the Government of Spain (FIS2020-TRANQI and Severo Ochoa CEX2019-000910-S),
Fundaci\'o Cellex, Fundaci\'o Mir-Puig, Generalitat de Catalunya (CERCA program),
and the EU NextGen Funds.}

\dataavailability{No new data were created or analysed in this study. Data sharing is not applicable to this article.}

\acknowledgments{The authors thank Guillermo Curr\'as-Lorenzo for pointing
out a flaw in an earlier version of the analysis in this manuscript.}

\conflictsofinterest{The authors declare no conflicts of interest.}

\begin{adjustwidth}{-\extralength}{0cm}

\reftitle{References}
\externalbibliography{yes}
\bibliography{practical}

\end{adjustwidth}

\end{document}